\documentclass[12pt]{iopart}
\usepackage{citesort}
\usepackage{siunitx}
\usepackage{graphicx}

\usepackage{amssymb,amsfonts} 

\usepackage{comment}
\usepackage{tikz}

\newcommand{\midarrow}{\tikz \draw[ -stealth] (0,0) -- +(.1,0);}
\newcommand{\midarrowrev}{\tikz \draw[stealth -] (0,0) -- +(.1,0);}
\newcommand{\middownarrow}{\tikz \draw[stealth -] (0,0) -- +(0,.1);}
\newcommand{\miduparrow}{\tikz \draw[-stealth ] (0,0) -- +(0,.1);}

\tikzset{->-/.style={decoration={
  markings,
  mark=at position .6 with {\arrow{stealth}}},postaction={decorate}}}

\tikzset{bareU/.style={decorate, decoration={snake}, draw = black}}

\newcommand{\Pch}{
\tikz {
\draw [fill = gray] (0.5,0) rectangle (0.75,1);
\draw (0,0)  --node {\midarrow}(0.5,0) ;
\draw (0,1) --node {\midarrow}(0.5,1) ;
\draw (0.75,0)  --node {\midarrow}(1.25,0) node[] at (1.25, -0.5) {$\dot P^{\Lambda}$};
\draw[dashed] (0.75,1) --node {\midarrow}(1.25,1) ;

\draw [fill = gray] (1.75,0) rectangle (2.,1);
\draw (1.25,0)  --node {\midarrow}(1.75,0) ;
\draw[dashed] (1.25,1) --node {\midarrow}(1.75,1) ;
\draw (2,0)  --node {\midarrow}(2.5,0);
\draw (2,1) --node {\midarrow}(2.5,1);
}}

\newcommand{\Cch}{
\tikz {
\draw [fill = gray] (0.5,0) rectangle (0.75,1);
\draw (0,0)  --node {\midarrowrev}(0.5,0) ;
\draw (0,1)  --node {\midarrow}(0.5,1) ;
\draw (0.75,0)  --node {\midarrowrev}(1.25,0) node[] at (1.25, -0.5) {$\dot C^{\Lambda}$};
\draw[dashed] (0.75,1) --node {\midarrow}(1.25,1) ;

\draw [fill = gray] (1.75,0) rectangle (2.,1);
\draw (1.25,0)  --node {\midarrowrev}(1.75,0) ;
\draw[dashed] (1.25,1) --node {\midarrow}(1.75,1) ;
\draw (2,0)  --node {\midarrowrev}(2.5,0);
\draw (2,1) --node {\midarrow}(2.5,1);
}}

\newcommand{\Dch}{
\tikz {
\draw [fill = gray] (0.5,0.5) rectangle (0.75,1);
\draw (0,1) --node {\midarrow}(0.5,1) ; 
\draw (0.75,1) --node {\midarrow}(1.25,1) ;
\draw (0.5,0.5) arc (90:180:0.25) node{\middownarrow} arc (180:270:0.25);
\draw[dashed] (0.75,0.5) arc (90:0:0.25) node{\miduparrow} arc (0:-90:0.25);
\draw [fill = gray] (0.5,0) rectangle (0.75,-0.5);
\draw (0.0,-0.5)  --node {\midarrow}(0.5,-0.5) ;
\draw (0.75,-0.5)  --node {\midarrow}(1.25,-0.5);

\node[] at(1.75,0.25){$+$};
\node[] at(4.62,0.25){$+$};

\draw [fill = gray] (3.1,1) rectangle (3.35,0.25);
\draw (2.25,1)  --node {\midarrow}(3.1,1) ; 
\draw (3.35,1) --node {\midarrow}(4.25,1) ;

\draw[->-] (2.75,-0.25) -- (3.1,0.25);
\draw[->-, dashed] (3.35,0.25) -- (3.75,-0.25);

\draw [fill = gray] (2.75,-0.5) rectangle (3.75,-0.25);
\draw (2.25,-0.5)  --node {\midarrow}(2.75,-0.5) ;
\draw (3.75,-0.5)  --node {\midarrow}(4.25,-0.5);

\draw [fill = gray] (5.5,1) rectangle (6.5,0.75);
\draw (5,1)  --node {\midarrow}(5.5,1);
\draw (6.5,1)  --node {\midarrow}(7,1);

\draw[->-] (5.5,0.75) -- (5.85,0.25);
\draw[->-, dashed] (6.1,0.25) -- (6.5,0.75);

\draw [fill = gray] (5.85,-0.5) rectangle (6.1,0.25);
\draw (5,-0.5) --node {\midarrow}(5.85,-0.5) ; 
\draw (6.1,-0.5) --node {\midarrow}(7,-0.5);
\draw [decorate,decoration={brace,amplitude=10pt,raise=4pt},yshift=0pt]
(7,-0.5) -- (0,-0.5) node [black,midway,yshift=-0.8cm] {$\dot D^{\Lambda}$};
}}

\newlength{\overwritelength}
\newlength{\minimumoverwritelength}
\newcommand{\overwrite}[3]{%
  \settowidth{\overwritelength}{$#1$}%
  \ifdim\overwritelength<\minimumoverwritelength%
    \setlength{\overwritelength}{\minimumoverwritelength}\fi%
  \stackrel
    {%
      \begin{minipage}{\overwritelength}%
        \color{#2}\centering\small #3\\%
        \rule{1pt}{9pt}%
      \end{minipage}}
    {\colorbox{#2!30}{\color{black}$\displaystyle#1$}}}

\usetikzlibrary{arrows,shapes,positioning}
\usetikzlibrary{decorations.markings}
\usetikzlibrary{decorations.pathmorphing}
\usetikzlibrary{decorations.pathreplacing}

\usetikzlibrary[arrows]
\tikzstyle{arrowstyle}=[scale=1]
\usepackage{tkz-euclide}

\usetikzlibrary{arrows,shapes,positioning}
\usetikzlibrary{decorations.markings}
\usetikzlibrary{decorations.pathmorphing}
\usetikzlibrary{decorations.pathreplacing}

\usetikzlibrary[arrows]
\usetikzlibrary{fadings,shapes.arrows,shadows}
\tikzstyle{arrowstyle}=[scale=1]
\usepackage{float}

\usepackage{url}
\usepackage{xurl}
\begin{document}

\title[Strong Boundary and Trap Potential Effects]{Strong Boundary and Trap Potential Effects on Emergent Physics in Ultra-Cold {Fermionic} Gases}

\author{J.B. Hauck$^1$, C. Honerkamp$^{2,3}$ and D.M. Kennes $^{1,3,4}$}\address{$^1$ Institut für Theorie der Statistischen Physik,  RWTH Aachen,  52056 Aachen, Germany and JARA - Fundamentals of Future Information Technology}\address{$^2$ Institute for Theoretical Solid State Physics, RWTH Aachen University, 52074 Aachen, Germany}\address{$^3$ JARA-FIT, Jülich Aachen Research Alliance - Fundamentals of Future Information Technology, Germany}\address{$^4$ Max Planck Institute for the Structure and Dynamics of Matter and Center for Free Electron Laser Science, 22761 Hamburg, Germany}
\ead{hauck@physik.rwth-aachen.de}
\vspace{10pt}
\begin{indented}
\item[]April 2021
\end{indented}

\begin{abstract}
The field of quantum simulations in ultra-cold atomic gases has been remarkably successful. In principle it allows for an exact treatment of a variety of highly relevant lattice models and their emergent phases of matter. But so far there is a lack in the theoretical literature concerning the systematic study of the effects of the trap potential as well as the finite size of the systems, as numerical studies of such non periodic, correlated fermionic lattices models are numerically demanding beyond one dimension. We use the recently introduced real-space truncated unity functional renormalization group to study these boundary and trap effects with a focus on their impact on the superconducting phase of the $2$D Hubbard model. We find that in the experiments not only lower temperatures need to be reached compared to current capabilities, but also system size and trap potential shape play a crucial role to simulate emergent phases of matter.
\end{abstract}

\submitto{\NJP}

\section{Introduction}
Ultra-cold atomic gases provide a powerful method for simulations of quantum many-body lattice systems~\cite{bloch_many-body_2008,giorgini_theory_2008,tarruell_quantum_2018}. Many models introduced in the context of solid-state physics, like the Hubbard model~\cite{tarruell_quantum_2018}, the Haldane model~\cite{jotzu_experimental_2014}, the bilayer Hubbard model~\cite{gall_competing_2021} or even quasicrystalline models~\cite{mace_quantum_2016}, can be investigated in laboratories without the restriction of numerical ambiguities or theoretical approximation errors. Beyond this, other exciting many-body phenomena that cannot be easily explored in solids could be realized in optical lattices as well{, see~\cite{gross_quantum_2017,goldman_topological_2016,bloch_many-body_2008,lewenstein_ultracold_2012} and references therein}, making them a very flexible tool to study quantum many-body physics.

Especially in the context of the fermionic Hubbard model in optical lattices, there have been great advances in recent years, including the observation of the Mott insulating phase~\cite{jordens_mott_2008,Schneider2008,cheuk_observation_2016} and first sign of antiferromagnetic correlations~\cite{mazurenko_cold-atom_2017,hart_observation_2015}. Despite these advances, in the context of correlated fermions state-of-the-art experiments still do not reach beyond the capabilities of exact numerical methods, like quantum Monte Carlo (QMC) or exact diagonalization, as all observed phases can also be investigated quantitatively with the available set of theoretical methods. This may change when the experiments move towards the exploration of $d$-wave superfluidity~\cite{hofstetter_high-temperature_2002}, where, e.g.,  QMC encounters the notorious sign problem and finite size scaling from exact diagonalization is hopeless. Despite the lack of quantitative studies of trap or finite size effects in this regard, the current hope is that if the experiments can reach lower temperatures, the superfluid phase will emerge naturally. 

{However, the effects of the trapping potential have always been an issue in fermionic~\cite{rigol_local_2003,rigol_quantum_2004,scarola_discerning_2009,chiesa_magnetism_2011,cone_optimized_2012,duarte_compressibility_2015,mendes-santos_size_2015,nigro_trap_2017,chanda_coexistence_2020} and bosonic~\cite{batrouni_mott_2002,wessel_quantum_2004,kato_quantum_2009,delande_compression_2009} optical lattices, even for the Mott-phase and high temperature properties of one and two dimensional Hubbard models. In this paper we will focus on fermionic setups. It is not expected that the observed effects are the same or even similar for bosons and fermions as their behavior is already very  different for non-interacting particles. Additionally, we concentrate on two-dimensional systems only.} In continuation of these previous results, we investigate the effects of the trapping potential, as well as the boundary conditions on the  Cooper pairing in a Hubbard model. This is of high experimental importance as one of the current goals is to measure the superfluid phase. A general challenge for Cooper pairing in finite-size systems can, e.g., be gleaned from the literature for superconductivity of ultra-small aluminium grains~\cite{von_delft_superconductivity_2001,von_delft_parity-affected_1996}, with a diameter of the order of $\propto \si{\nano\metre}$. There, it was worked out clearly how the usual superconducting description is valid as long as the energy gap due to pairing is larger than the energy spacing due to the confinement but breaks down if the latter scale is the larger one. A similar crossover is expected to happen also in a Hubbard model. Even from these basic considerations, the question arises whether there is a minimal number of sites required to unveil superconducting hallmarks. What might add in the case of $d$-wave pairing is the question, if the generation of the attractive interaction, at weaker coupling mainly by spin fluctuations, is somehow altered by the trap confinement.
Finally, it is an open question how the non-trivial $d$-wave ordering pattern will be influenced by the trapping potential. In the antiferromagnetic case the trap was not observed to be a significant perturbation~\cite{mazurenko_cold-atom_2017}. In fact the data compared well to Quantum Monte Carlo results obtained for a square lattice with periodic boundary conditions. 
Here we want to examine whether this holds true also for the d-wave superconducting phase in a square lattice Hubbard model. 
To extrapolate to the bulk case (e.g.,~to connect to real materials), one should also understand how closely the finite-size set-up resembles the situation in the thermodynamic limit. Hence, we will also examine at which size the trapped system approximates the bulk case to high fidelity.

\section{Setup}
\begin{figure}[!htbp]
\centering
\includegraphics[width = 0.32\linewidth]{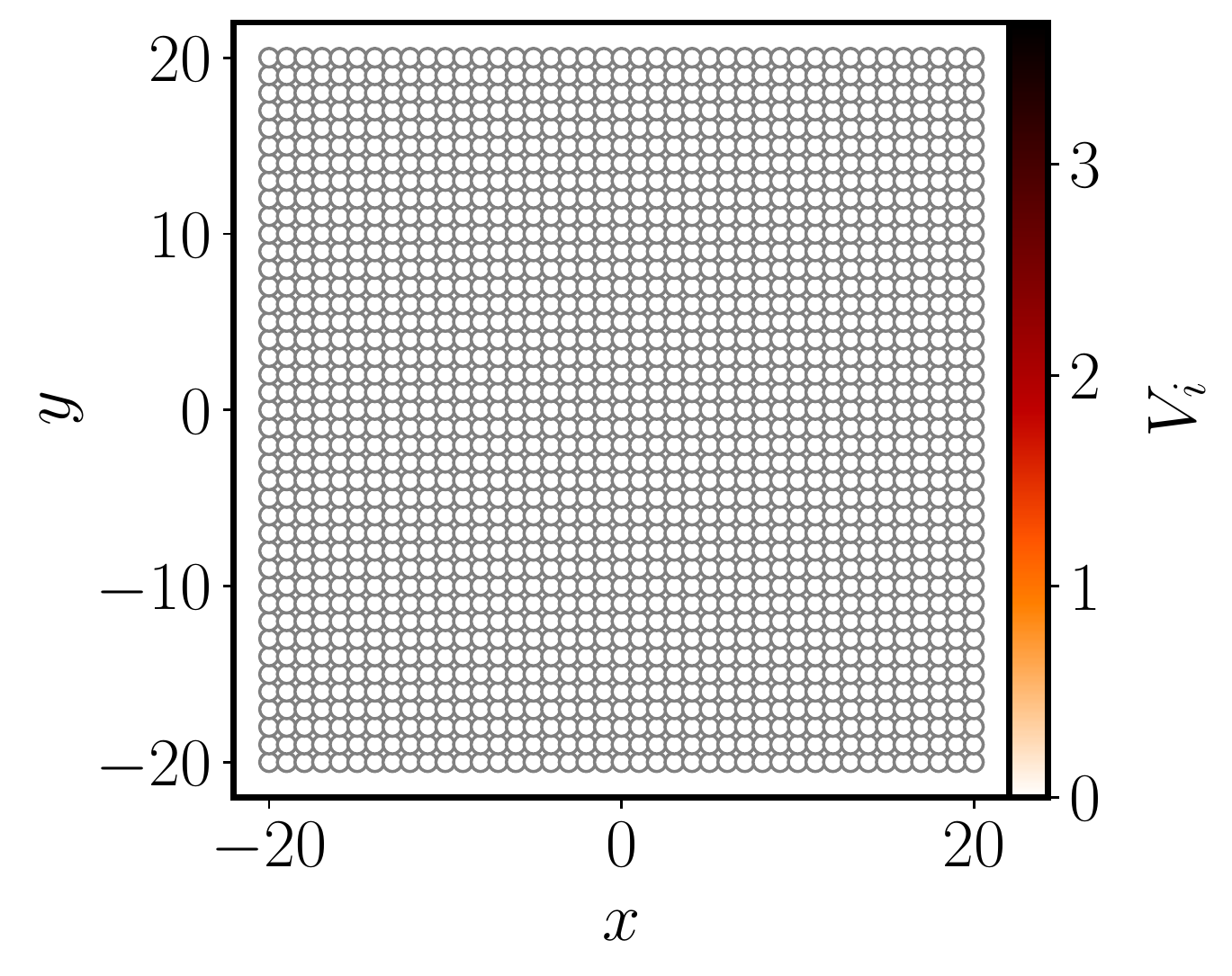}
\includegraphics[width = 0.32\linewidth]{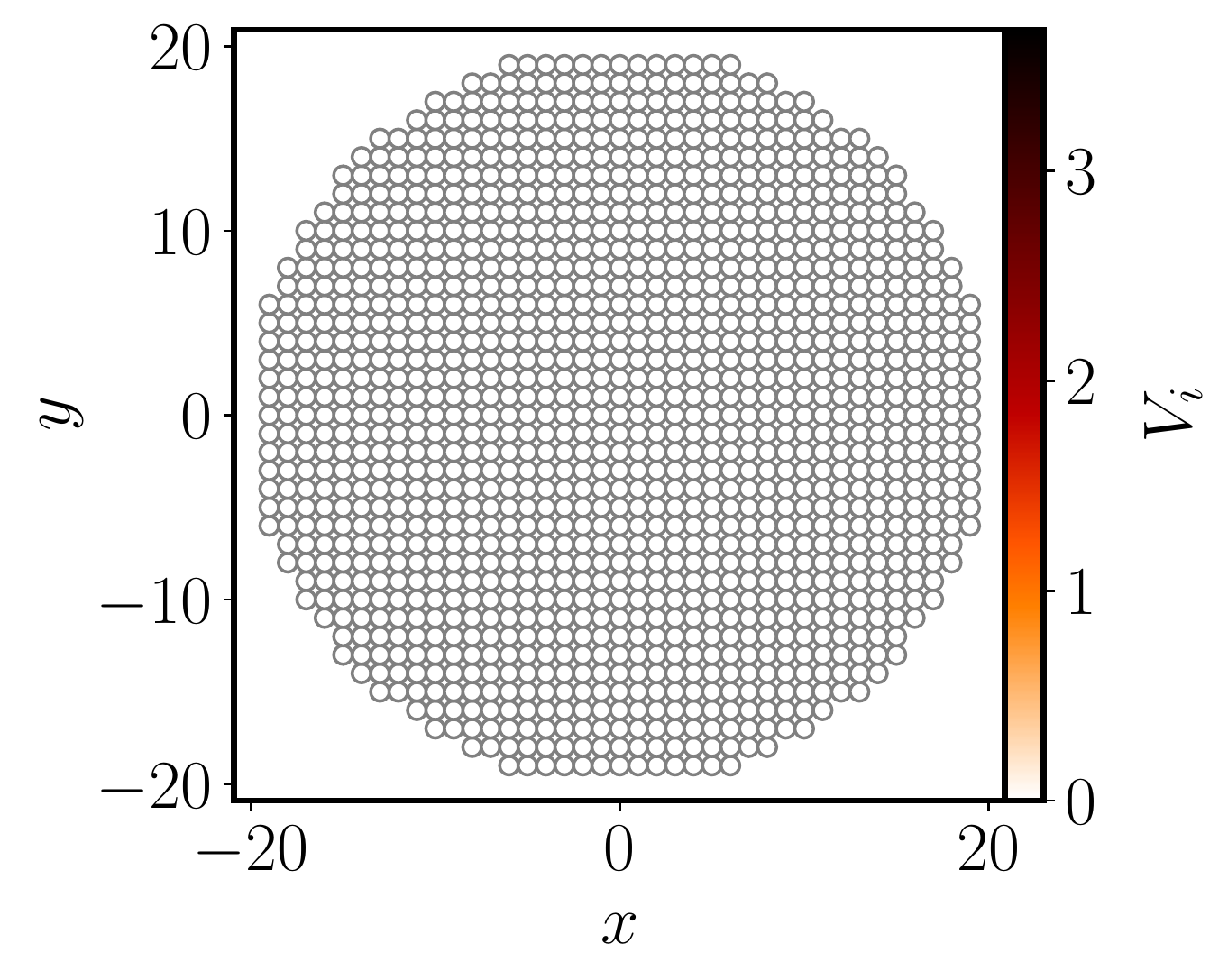}
\includegraphics[width = 0.32\linewidth]{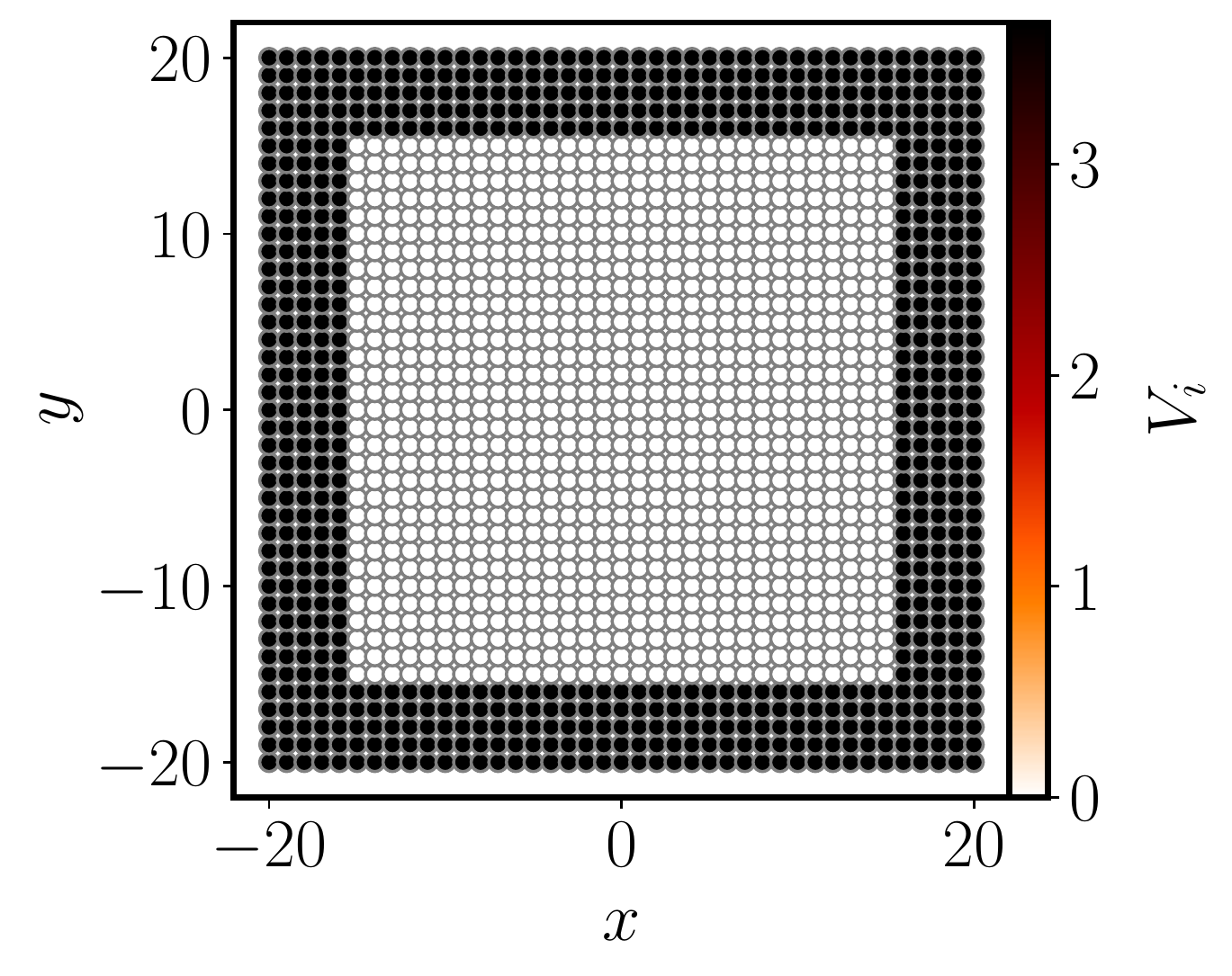}
\includegraphics[width = 0.32\linewidth]{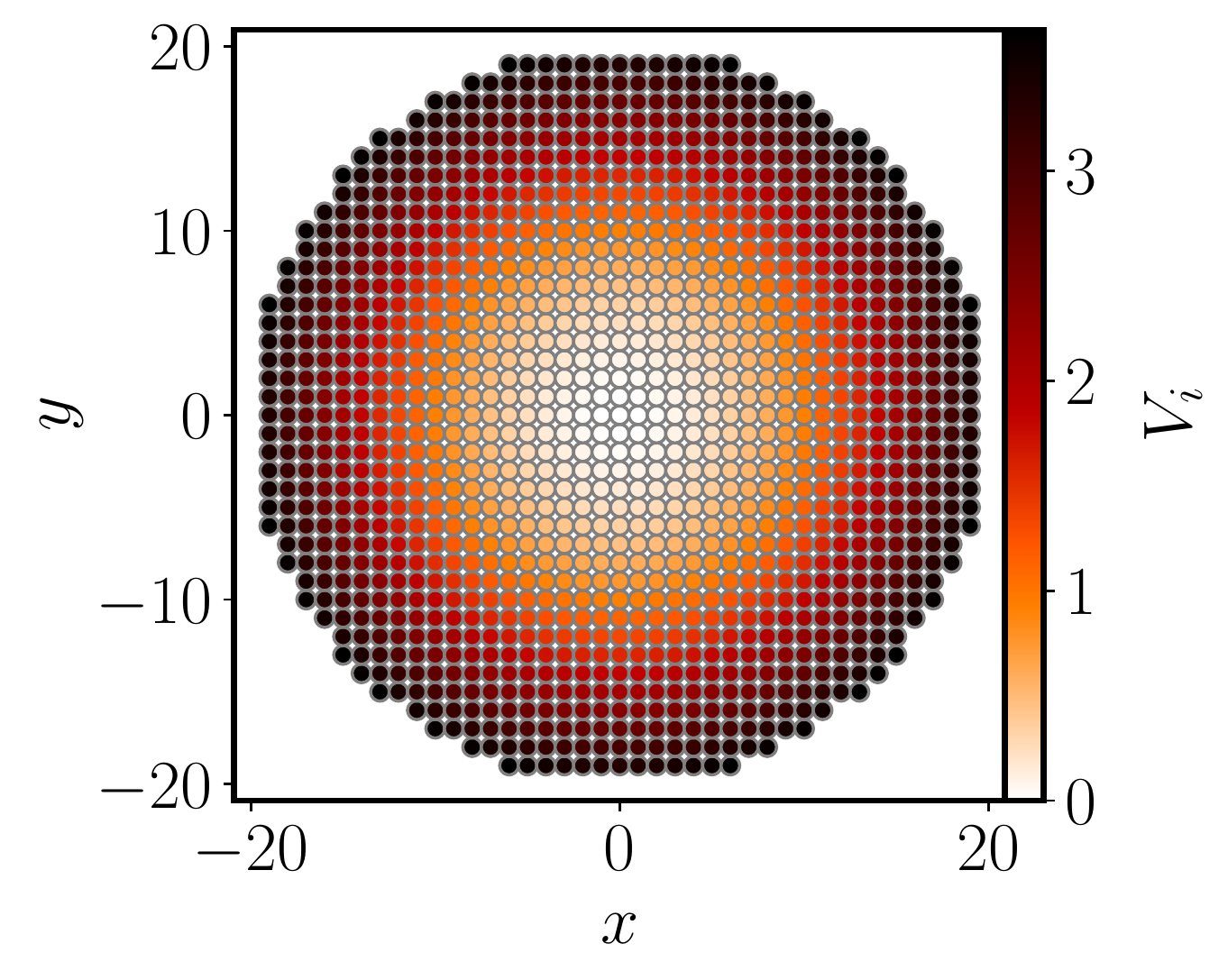}
\includegraphics[width = 0.32\linewidth]{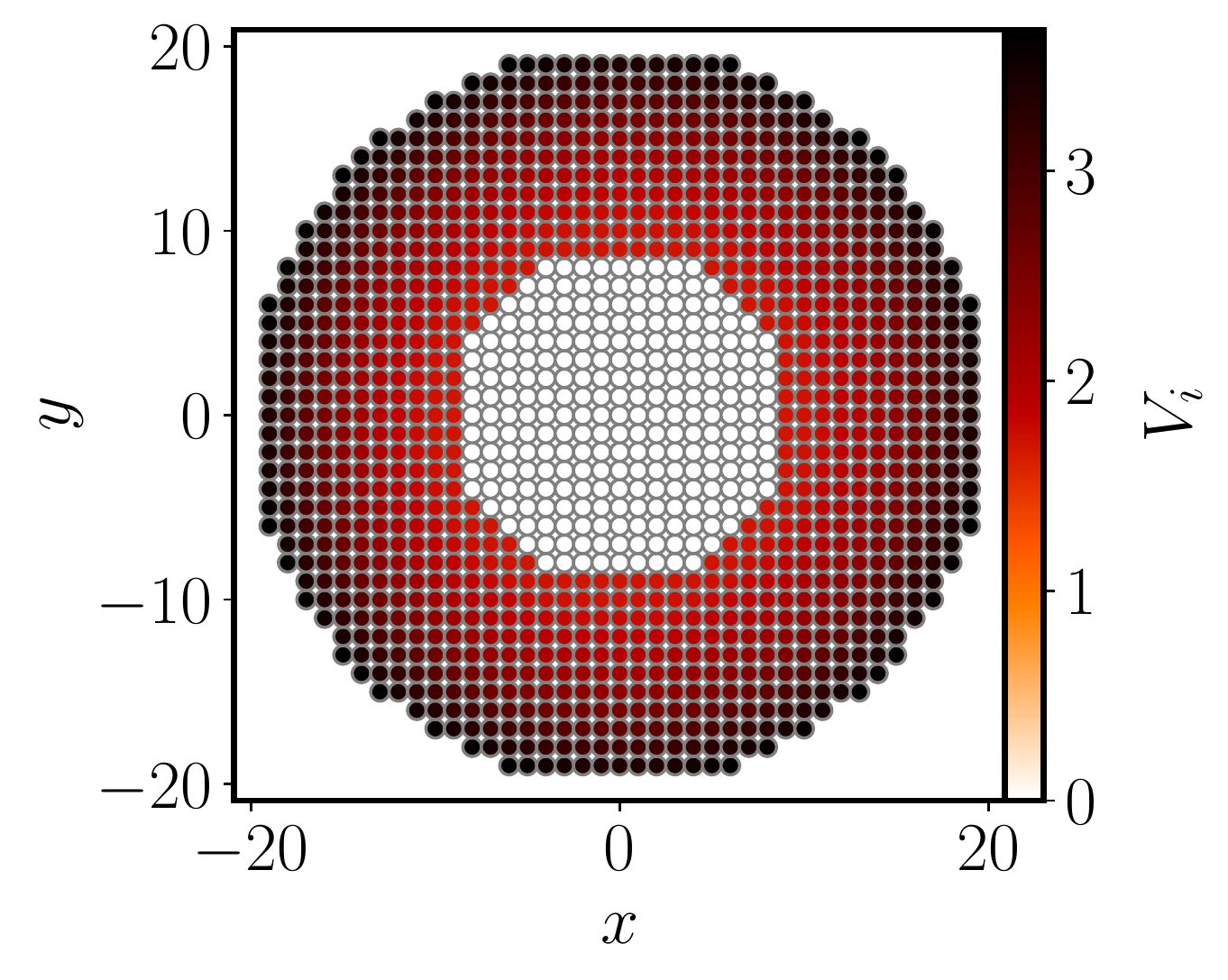}
\caption{Schematic visualization of the five different geometries and trapping combinations considered here. In the upper left, the simple square lattice with open boundary conditions is visualized and in the upper middle plot, the circular open boundary conditions case is shown. The upper right plot shows the combination of open boundary conditions with a finite height trapping. The lower plots visualize the two different choices of potentials applied in the circular open boundary conditions case, where the left is the simple quadratic trap, and the right approximates the experimental trap. The color scale gives the values of the trapping potential $V_i$.}
\label{fig::trap_pot}
\end{figure}

We will simulate five different setups, all are different realizations of the square lattice Hubbard model with different open boundary conditions. They differ in the lattice confining geometry, which is either a square box or a circle, and trapping potentials, the different setups are visualized in Fig.~\ref{fig::trap_pot}. The first setup investigated is a square lattice Hubbard model with open boundary conditions and without any confining potentials. The second one is a square lattice Hubbard model with circular open boundary conditions and without any confining potential. These two cases are investigated to distinguish between the effects introduced by the open boundary conditions applied in all cases and the ones introduced by the different lattice shapes. The third case is a square lattice Hubbard model with open boundary conditions in which a second box potential with finite height is embedded. Thereby the open boundary conditions are smoothed. As the fourth setup we consider a square lattice Hubbard model with circular open boundary conditions, in which a quadratic trap is inserted, with $V_{i}^{trap} = a_{trap}r_i^2$, where $r_i$ is the radial distance of site $i$ to the center of the lattice and $a_{trap}$ gives the curvature of the trapping potential. The fifth and last setup is the same as the fourth with a different potential: Instead of the simple quadratic one, we reconstruct a potential used in a recent experiment by Mazurenko et.~al.~\cite{mazurenko_cold-atom_2017}. There the trap consists of a Gaussian potential created by a digital micro-mirror device (DMD) with an additional circular quadratic potential, whose superposition creates a nearly flat disc at the center. For simplicity we approximate it by a piece-wise definition, see Eq.~(\ref{Eq:picewise_trap}). 
The {square lattice} Hamiltonian in second quantization for all cases reads
\begin{equation}
H = -\sum_{i,j, \sigma} [t_{ij} + (\mu+V_i^{trap}) \delta_{i,j}]c^{\dagger}_{i,\sigma} c_{j,\sigma} + \frac{1}{2}\sum_{i,j,\sigma,\sigma'} U_{i,j} n_{i, \sigma} n_{j, \sigma'},
\label{Eq:Qsim_hub}
\end{equation} 
with the operators $c_{i,\sigma}^{(\dagger )}$ annihilating (creating) an  electron on site $i$ with spin $\sigma$.
For the hopping amplitudes we set $t_{ij} = t$ if $i$ and $j$ are nearest-neighbors and $t_{ij} = t'$ if $i$ and $j$ are next-nearest-neighbors, { where we define nearest and next-nearest neighbors in terms of the distance on the underlying lattice as the closest and second closest sites}. All other terms are set to zero. For simplicity we choose $t=1$ and measure all quantities in units of $t$. Distances are measured in units of the lattice spacing $a$ which we set to $1$. 
{We choose onsite interactions $U_{ij} = U\delta_{i,j}$ and fix the 'Van-Hove condition' $\mu = 4t'$. The Van Hove condition ensures that our setup is tuned to a van Hove singularity~\cite{Hove_occurence_1953} of the thermodynamic limit Hubbard model, at which the gradient of the energy dispersion is flat and therefore the density of states is large. We neglect density dependent hoppings as it scales with $U$, which is comparably small in our setup, and $n_i$, which is always below half filling, therefore its effect should be negligible~\cite{meinert_quantum_2013,dutta_non-standard_2015,amadon_metallic_1996}.}
We use an extended Hubbard model as recent numerical studies \cite{qin_absence_2020,zheng_stripe_2017,leblanc_solutions_2015} suggest that in the relevant $U$ parameter range for cuprates there is no superconductivity in the pure ($t'=0$) Hubbard model.  

\section{Method} 
\label{sec:FRG}
We employ the recently introduced $2$D real-space functional renormalization group (real-space TUFRG)~\cite{hauck_electronic_2020}, which can be used to study models with broken translational symmetry. It is based on the standard one-particle-irreducible FRG~\cite{metzner_functional_2012,lichtenstein_high-performance_2017}, in a level-2 truncation. Additionally, we will neglect the frequency dependence of the vertex, as well as the self-energy. With this we obtain a  method, which scales relatively well and is able to predict finite-size precursors to phase transitions and their respective orderings. The main idea of FRG is the introduction of a cutoff in the bare propagator, enabling an interpolation between a solvable theory and the full solution of the model. Variations with respect to the cutoff parameter generate an infinite set of coupled differential flow equations for the Taylor coefficients of the effective action. These equations must be truncated in order to become numerically tractable, which is well defined in the case of $\frac{U}{W}<1$, with $W$ being the bandwidth of the model, then only the first few terms in the expansion are relevant and a truncation at the level of keeping just the self-energy and the effective interaction is justified. 
Explicitly the cutoff is introduced in the bare Green's function by
\begin{equation}
    G_{0}(\omega_n) = \frac{1}{i\omega_n-H} \longrightarrow G_{0}^\Lambda(\omega_n) = \frac{R(\Lambda)}{i\omega_n-H}
    \label{Eq:G}
\end{equation}
with $R(\Lambda)$ being the cutoff function obeying $R(0) = 1$ and $\lim\limits_{\Lambda \rightarrow \infty} R(\Lambda) = 0$ in our convention. This reduces the action in the limit $\lambda\rightarrow \infty$ to the non-interacting, exactly solvable system. By integrating the coupled flow equations, an interpolation between this solution of the non-interacting problem and the approximate full solution is calculated. The variations w.r.t~the cutoff parameter generate the single scale propagator defined as 
\begin{equation}
    S^\Lambda(\omega_n) = \frac{\dot{R}(\Lambda)}{i\omega_n-H}
    \label{Eq:singscale}
\end{equation}
in our approximation.

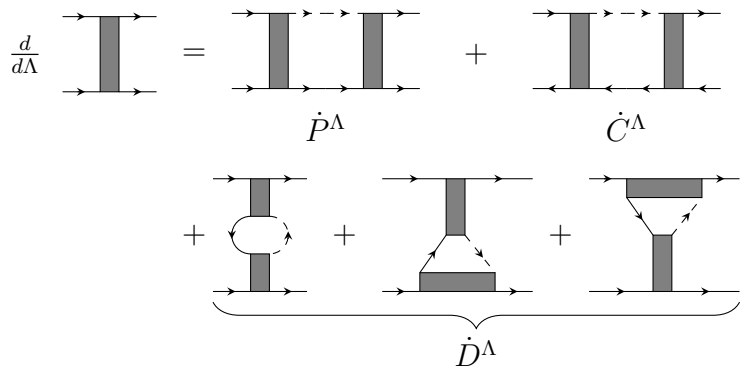
\begin{figure}
    \begin{center}

    {
\scalebox{1}
{

\begin{tikzpicture}
\node[] at(-1,0.5){$\frac{d}{d\Lambda}$};
\draw [fill = gray] (0,0) rectangle (0.25,1);
\draw (-0.5,0)  --node {\midarrow}(0,0) ;
\draw (-0.5,1)  --node {\midarrow}(0,1) ;
\draw (0.25,0)  --node {\midarrow}(0.75,0);
\draw (0.25,1) --node {\midarrow}(0.75,1);

\node[] at(1.25,0.5){$=$};

\node[] at(3,0.2){\Pch};

\node[] at(5,0.5){$+$};

\node[] at(7,0.2){\Cch};

\node[] at(1.25,-1.9){$+$};

\node[] at(5,-2.4){\Dch};

\end{tikzpicture}}
}

\end{center}
    \caption{Diagrammatic representation of the FRG flow equation for the one-particle irreducible interaction vertex in the level-II truncation, adapted from~\cite{metzner_functional_2012}. The first term on the right-hand side, annotated with $\dot P^\Lambda$, represents the flow of the particle-particle channel $P^\Lambda$, while the other terms generate the flow of $C^\Lambda$ and $D^\Lambda$. The dashed internal lines represent a single scale propagator, see Eq.~(\ref{Eq:singscale}), and the solid lines represent a propagator, see Eq.~(\ref{Eq:G}). The arrows indicate the direction of the propagators and fix the order of their arguments. The grey boxes represent the full two-particle interaction. A second class of diagrams with the scale-derivative on the other internal line contributes as well but is now drawn for simplicity. }
\label{Fig:oneloopdiags}
\end{figure}

The scale-dependent two-particle vertex can be separated in three different channels according to the three different fermionic bilinears or three different diagram types as shown in Fig.~\ref{Fig:oneloopdiags}. Each of those bilinears is directly associated to a bosonic field of an effective Hamiltonian \footnote{In fact these bilinears are interpreted most easily for the particle-particle ($P$) and the crossed particle-hole ($C$=spin) channels, the direct particle-hole channel ($D$) is more complex as it contains information on the $C$-channel. The physical relevant quantity is the charge channels given by $V_{ch} = 2D-C$, which removes the redundant information.}. The particle-particle channel ($P^\Lambda$) is associated with the superconducting pairing interactions and therefore a divergence of this channel at total momentum zero indicates a tendency towards superconductivity. The direct particle-hole channel ($D^\Lambda$) is associated to charge fluctuations and thus indicates, among others, general charge ordering. The crossed particle-hole channel ($C^\Lambda$) can be associated with effective spin-spin interactions, therefore its divergence is an indicator of magnetic ordering. 
In typical flows, the effective interaction undergoes a flow to strong coupling, i.e. diverges at a certain finite RG scale. When this occurs, we stop the flow as soon as an eigenvalue of one of the three channels surpasses a threshold value. The corresponding RG scale is called the critical scale $\Lambda_c$, which defines an energy scale of the breakdown of the weakly interacting behavior and can be seen as an estimate of an ordering temperature $T_c$. The leading eigenvalue of the diverging channel gives, by its eigenvector, the expected ordering pattern up to an unknown prefactor. A more precise determination of the order would follow from the gap equations for each of the channels. In practice we will normalize the shown ordering pattern from the leading eigenvector such that their maximal absolute value is $1$.  As we consider finite-size systems here, the interpretation of the flows to strong coupling as phase transitions is not strictly valid, as neither a real discontinuity occurs, nor any correlation length truly diverges. Nevertheless, it can be expected that the finite-size system with a strongly enhanced interaction of a specific type will locally resemble a infinite system with a truly diverging effective interaction of the same type. Thus, we use the same vocabulary as for the infinite system and discuss different phases of the finite system as regimes with qualitatively different flows to strong coupling. 

{
As we do not include self-energy effects, we cannot incorporate possible gap openings, which would allow us to continue the flow into the symmetry broken region~\cite{Salmhofer_2004,Eberlein_2014,Wang_2014,Maier_2014}. Therefore, lower lying divergences in other locations of the trap cannot be resolved within our approach. 
In the case that there are multiple eigenvalues diverging at once, we can access the sub-leading eigenvalues and could perform a post-processing mean-field decoupling in order to obtain the expected gap pattern, thereby, resolving cases in which the dominant eigenvector does not contribute to the order parameter. The results presented here showed in all cases a clear dominant eigenvector instead of divergences of multiple ones. Therefore the above described procedure was never needed.
}

\begin{figure}
    \begin{center}

    {
\scalebox{1}
{

\begin{tikzpicture}
\node[] at(-1.3,0.5){$U_{i,i}=$};
\draw [bareU] (0,0) --(0,1);
\node[below] at (0.0,0) {$i$};
\node[above] at (0.0,1) {$i$};

\draw (-0.5,0)  --node {\midarrow}(0.5,0);
\draw (-0.5,1) --node {\midarrow}(0.5,1);

\node[] at(1,0.5){$\rightarrow$};

\node[] at(2,0.5){$P^{(2)} =$};
\draw [bareU] (3.5,0) --(3.5,1);
\node[below] at (3.5,0) {$i$};
\node[above] at (3.5,1) {$i$};
\draw (3,0)  --node {\midarrow}(5.5,0);
\draw (3,1) --node {\midarrow}(5.5,1);

\draw [bareU] (5,0) --(5,1);
\node[below] at (5,0) {$j$};
\node[above] at (5,1) {$j$};
\end{tikzpicture}}
}

\end{center}
    \caption{Generation of new index dependencies during the flow. A density-density interaction $U_{i,i}n_{i,s}n_{i,s'}$ receives corrections in the $P$-channel that can be interpreted as an incoming particle pair $c_{i,s}c_{i,s'}$ scattering to an outgoing pair $c_{j,s}c_{j,s'}$. }
\label{Fig:example}
\end{figure}
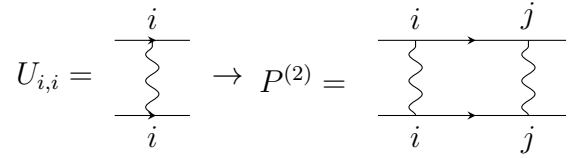

Next, we discuss the spatial dependence of the effective interactions and the respective channels $P$, $C$ and $D$. In principle, each channel can depend on four site indices. Yet, the initial condition in Eq.~(\ref{Eq:Qsim_hub}) is an onsite interaction, i.e. where all four site indices have to be the same\footnote{This argumentation generalises to non-onsite interactions but we focus on the simplest case for the sake of simplicity}. Looking at the second-order correction e.~g.~in the $P$-channel as in Fig.~\ref{Fig:example}, we observe that this adds a non-local, potentially long-ranged, 'bi-local' correction that depends on the joint index of the incoming lines $i$ and the outgoing lines $j$. Like in the $P$-channel, these RPA-like contributions create a main dependence on different bi-local pairs of indices in each channel. Contributions beyond these bi-local indices are only generated by the feedback in between the channels and, thus, come at a higher order in the interaction. The further apart from the native bi-local indices, the higher is the interaction order~\cite{markhof_detecting_2018}. Therefore, these terms can be neglected in the spirit of the perturbative motivation of the FRG. The bi-local contributions of each channel on the other hand have to be captured more rigorously, as they are equivalent to the sharp momentum structures that builds up in the translationally invariant case~\cite{husemann_2009,wang_smfrg,lichtenstein_high-performance_2017}.

Formally, this specific structure of the effective interactions can be exploited by an expansion of the subleading dependencies in form factors, which form a complete orthonormal basis set, according to
\begin{eqnarray}
\sum_{i} f_{b_k}(i)  f^{*}_{b_{k'}} (i) = \delta_{b_k,b_{k'}}, \quad
\sum_{b_k} f_{b_k}(i)f^{*}_{b_k}(i') = \delta_{i,i'}.
\label{Eq::ONB}
\end{eqnarray}
The three channels can then be build up from form factors capturing the internal structure of the bilinears that in turn can mutually interact on longer distances 
\begin{eqnarray}
\hat{P}[\Gamma]_{i,j}^{b_i,b_j}  = \sum_{k,l} \Gamma(i,k;j,l)f_{b_i}(k)f^{*}_{b_j}(l), \\* 
\hat{C}[\Gamma]_{i,j}^{b_i,b_j}  = \sum_{k,l} \Gamma(i,k;j,l)f_{b_i}(l)f^{*}_{b_j}(k), \\* 
\hat{D}[\Gamma]_{i,l}^{b_i,b_l}  = \sum_{k,l} \Gamma(i,k;j,l)f_{b_i}(j)f^{*}_{b_l}(k). 
\label{Eq::project1}
\end{eqnarray}
With this at hand we can rewrite the flow equations by an insertion of form-factor unit matrices.
The main idea of the truncated unity approach is to restrict the number of form-factors in these unities to much less than the lattice sites~\cite{lichtenstein_high-performance_2017,eckhardt_truncated_2020,hauck_electronic_2020} 
\begin{equation}
    \sum_{b_k \in L} f_{b_k}(i)  f^{*}_{b_k} (i) = \sum_{b_k} \delta_{b_k}^L f_{b_k}(i)  f^{*}_{b_k}(i) \approx 1,
\end{equation}
where $L$ is the set of all allowed bonds and analogously $\delta_{b_k}^L=1$ if $b_k \in L$ with bond length up to $b_{\mathrm{max}}$.

In the form-factor space the flow equations can be rewritten in terms of highly efficient block-matrix products. Due to the truncation, the full vertex cannot be recovered exactly but instead projections must be introduced in order to reconstruct the full vertex approximately for the cross-channel feedback. These projections can be found for example in Ref.~\cite{hauck_electronic_2020,markhof_detecting_2018,weidinger_functional_2017}.
Instead of including the full spin dependence we will assume an $SU(2)$ invariant problem. This is by no means necessary but simplifies the presentation. The diagrammatic flow equation for the interaction vertex, for such a $SU(2)$ invariant model, are displayed in Fig.~\ref{Fig:oneloopdiags}.

These diagrams translate to
\begin{eqnarray}
    \frac{d}{d\Lambda} P^{\Lambda} = -\hat{P}[\Gamma]^\Lambda \cdot \dot{\chi}_{pp} \cdot \hat{P}[\Gamma]^\Lambda, \nonumber\\*
    \frac{d}{d\Lambda} C^{\Lambda} = -\hat{C}[\Gamma]^\Lambda \cdot \dot{\chi}_{ph} \cdot \hat{C}[\Gamma]^\Lambda, \nonumber\\*
    \frac{d}{d\Lambda} D^{\Lambda} = 2\hat{D}[\Gamma]^\Lambda \cdot \dot{\chi}_{ph} \cdot \label{equ:flow_equations} \hat{D}[\Gamma]^\Lambda\\* - \hat{C}[\Gamma]^\Lambda \cdot \dot{\chi}_{ph} \cdot \nonumber \hat{D}[\Gamma]^\Lambda \\* -\hat{D}[\Gamma]^\Lambda \cdot \dot{\chi}_{ph} \cdot \nonumber \hat{C}[\Gamma]^\Lambda,
\end{eqnarray}
with the particle-hole bubble $\dot{\chi}_{ph}$ and the particle-particle bubble $\dot{\chi}_{pp}$ defined as 

\begin{eqnarray}
    \dot{\chi}_{ph  (i,j)}^{b_i,b_j} = 2\sum_{\omega>0} \Re{\left(G^{\Lambda}(\omega)_{i,j}S^{\Lambda}(\omega)_{j+b_j,i+b_i} + G \leftrightarrow S\right)} \\
    \dot{\chi}_{pp  (i,j)}^{b_i,b_j} = 2\sum_{\omega>0} \Re{\left(G^{\Lambda}(\omega)_{i,j}S^{\Lambda}(-\omega)_{i+b_i,j+b_j} + G \leftrightarrow S\right)}.
\end{eqnarray}

For the discussion of the results we will often refer to the $s$-wave and $d_{x^2-y^2}$-wave component of the leading eigenvector, which we define as follows: The leading eigenvector of a channel $I_{i,j}^{b_i,b_j}$ depends on a single site and a bond index, thus it can be written as $v_i^{b_i}$. The $s$-wave, or onsite component is in this notation defined as $v_i^{b_0}$, where $b_0$ refers to the bond with zero length. The definition of the $d_{x^2-y^2}$ component follows analogously to momentum space~\cite{lichtenstein_high-performance_2017}, we enumerate next-nearest-neighbor bonds from $1$ to $4$ in a clockwise fashion starting from the up pointing bond, then we define $v_{i}^{d_{x^2-y^2}} = v_i^{b_1}-v_i^{b_2}+v_i^{b_3}-v_i^{b_4}$. 

\subsection{Numerical Implementation}

For the integration, an adaptive fifth order Runge-Kutta scheme~\cite{dormand_family_1980} of the odeint library~\cite{ahnert_odeint_2011} is used. The matrix products on the right hand side of Eq.~(\ref{equ:flow_equations}), which are the numerically most demanding parts, are performed on GPU's. To reduce the number of Matsubara frequencies needed for the calculation of the propagator, we use the Pad\'e-approximation of the Fermi-function \cite{han_analytic_2017,ozaki_continued_2007}. The results are then verified using a heuristic tangent spacing scheme. Both yield fast convergence of the Bubble terms tested against the analytical result. In the calculations we incorporate $300$ frequencies, which is equivalent to an error of the bubble calculation of about $10^{-8}$ at the lowest used temperature ($T=10^{-3}$) in comparison to the analytic result. 
We use a mapping of the infinite integration range to a finite one $\Lambda = \frac{1-x}{x}$ to accelerate the integration~\cite{weidinger_functional_2017}. The integration is started from a minimum $x = 10^{-3}$ as smaller starting values yield no significant difference in the final results. To reduce the load imbalance between the channels we perform a completion of the square for the $D$-channel. Thereby we only need to calculate a single matrix product, as the missing part for the completion is $\frac{dC^{\Lambda}}{d\Lambda}$ which needs to be calculated anyway. 
As a regulator at finite temperature we choose so-called $\Omega$-cutoff, first introduced by Husemann et.~al.~\cite{husemann_efficient_2009}, given by
\begin{equation}
    R(\Lambda,\omega) = \frac{\omega^2}{\omega^2+\Lambda^2}   \Rightarrow \dot{R}(\Lambda,\omega) = \frac{-2\Lambda\omega^2}{(\omega^2+\Lambda^2)^2}.
\end{equation}
It has the advantage of a real IR-divergence regularization but is numerically more challenging as for example the interaction cutoff~\cite{honerkamp_interaction_2004}. Additionally, no numerical instabilities arise if the Hamiltonian has a zero eigenvalue. At zero temperature we employ a sharp cutoff, which drastically simplifies the bubble calculations~\cite{markhof_detecting_2018,klebl_functional_2020}.
In total, the flow equations scale like $ \mathcal{O}(N^3\cdot \bar{N_b}^{3})$ ($N$ is the number of orbitals and $\bar{N_b}$ is the average number of bonds per site) due to the matrix products, the bubble calculation scales like $ \mathcal{O}(N_f\cdot N^2\cdot \bar{N_b}^{2})$ ($N_f$ is the number of Matsubara frequencies included in the summation). Upon including the self-energy as well as the frequency dependence of the vertex (in the single channel coupling or ECLA sense~\cite{weidinger_functional_2017,reckling_approximating_2018}) the error scales with $U^3$, this scaling has been checked for the 1D-Hubbard model comparing to exact diagonalization. 
Additionally, we verified that our implementation is sufficiently efficient to reach the  thermodynamic limit of the $2$D-Hubbard model, where the known FRG results are reproduced.

\section{Results}
\subsection{Open boundary conditions - finite size effects}
We start by an analysis of the finite size effects in a square lattice Hubbard model with open boundary conditions and without any confining potentials. The Van-Hove condition is fixed by setting $\mu = 4t'$ and we vary $t'$ at different system sizes to find at which size the critical scales, as well as the occurring phases, are converged. 

\begin{figure}[!htbp]
    \centering
    \includegraphics[width = 0.49\columnwidth]{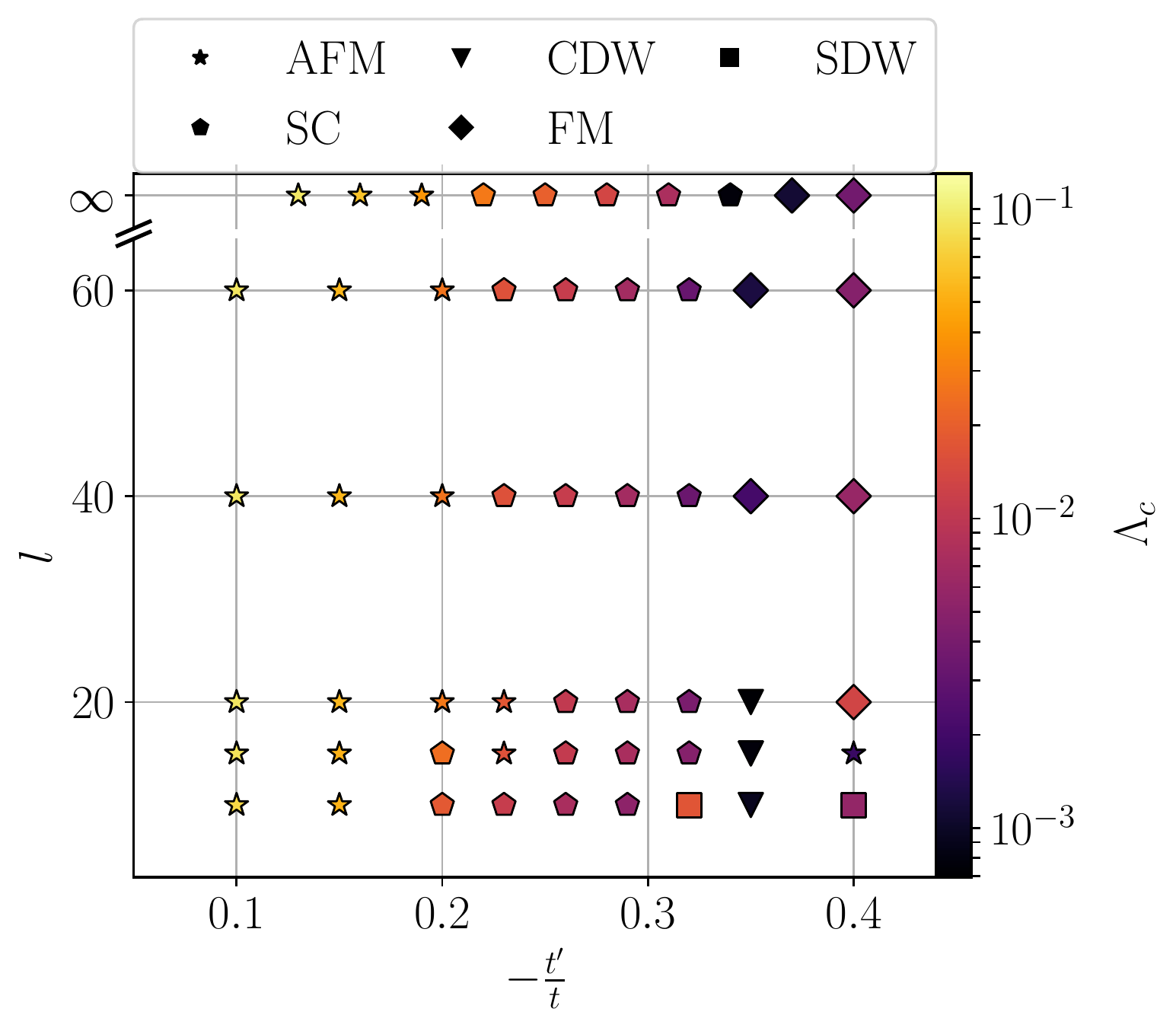}
    \includegraphics[width = 0.49\columnwidth]{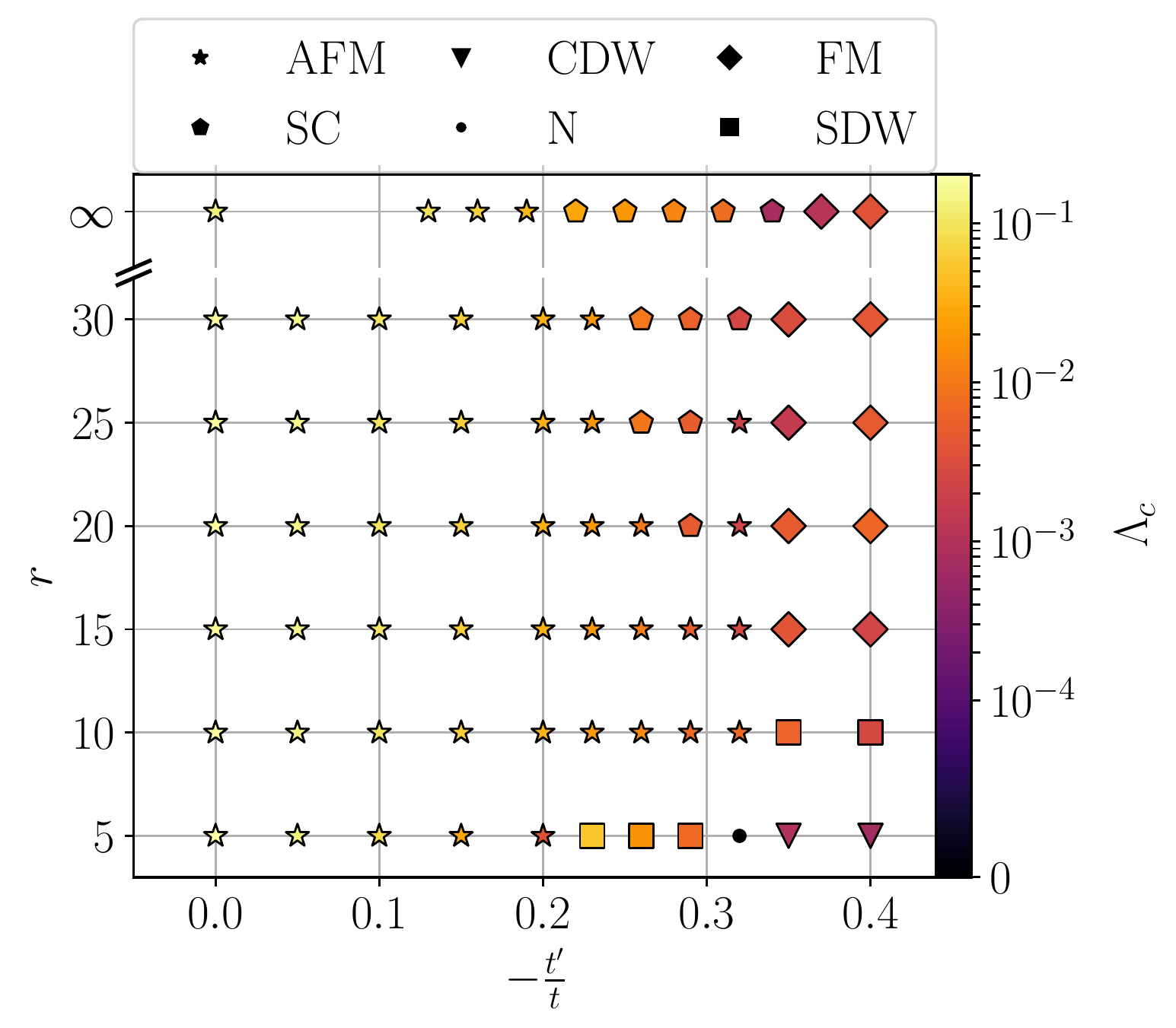}
    \caption{The phase diagram of the square lattice Hubbard model with open boundary conditions and without any confining potentials is shown on the left and the phase diagram of a square lattice Hubbard model with circular open boundary conditions and without any confining potential is shown on the right. Simulations are performed at van Hove filling at varying system sizes and next-nearest-neighbor hopping $\frac{t'}{t}$. We choose $U=3$, $T=10^{-3}$ and include nearest-neighbor correlations. The critical scales $\Lambda_c$ are encoded in the color scheme, whereas the type of the diverging phase is given by the shape of the data point. The thermodynamic limit results are given as a reference in the upper part of the plot. We encounter antiferromagnetism (AFM), charge density waves (CDW), general spin density waves (SDW), superconductivity (SC), ferromagnetism (FM) and no divergence at low scales (N).
    }  \label{fig:squarefs}
\end{figure}

The finite size effects turn out to be rather smooth except for the smallest linear size $l=10$ corresponding to $100$ lattice sites, see the left plot in Fig.~\ref{fig:squarefs}. The critical scales do vary only very slightly close to half filling. We find a superconducting phase even for the smallest system. The superconducting transition gets shifted to higher values of $-\frac{t'}{t}$ for larger systems, the largest system investigated shows a phase diagram roughly matching the thermodynamic limit case. It should be noted that the occurring superconducting phase is not a pure $d$-wave divergence, as one would obtain with periodic boundary conditions, but instead we have mixing between $s$- and $d$-wave bond pairing at all sizes at the boundary.
 Such an $s$- and $d$-mixing behavior has been a topic at the $[110]$ surface of square lattice systems (e.g.~\cite{honerkamp_instabilities_2000}), but here it is more profound as the onsite $s$-wave coupling is not put in à priori, but generated by the RG flow. Note that the initial interaction is a repulsive onsite interaction, but the RG flow generates, as a seemingly diverging tendency, an interaction between pair bilinears that include noticeable $s$-wave onsite terms, next to dominant nearest-neighbor pairs with $d$-wave pattern. Currently it is not clear which conditions, besides spatial inhomogeneity, are required to obtain this accompanying $s$-wave component. In our approach, the phase between the two components is fixed to unity as they arise in a single eigenvector.  The phase could change in the gap equation as this is a non-linear relation that could alter the onsite or bond pairings independently. Here, we will stick to the notion of $d+s$ mixing without a phase, indicating the dominant $d$-character.
For a proper verification of this finding, the possible gap opening should be investigated along the flow, which is numerically very demanding and, thus, is not done in this paper. The magnitude of the $s$-wave component relative to the one of the $d$-wave component decreases with increasing lattice size. In the largest system, the $s$-wave component has smaller amplitude than the $d$-wave component. It has additionally a stronger site dependence, i.~e.~, it shows its highest values at the boundaries. 
The superconducting phase does not fully resemble the thermodynamic limit case due to the visibility of boundary effects in the whole bulk even for $l=60$. {In summary, we observe a clear tendency towards an increased relevance of finite size effects with increasing next-nearest neighbor hopping.  Additionally, the superconducting and ferromagnetic ordering vectors we found to deviate substantially from the thermodynamic limit.}

For the experiments, round trapping potentials are easier to generate than {sharp} square trapping potentials~{\cite{mazurenko_cold-atom_2017,gall_competing_2021,hart_observation_2015}}. Therefore, the second setup under investigation is a square lattice Hubbard model with circular open boundary conditions and without any confining potential, which can be seen as a prototypical model for a sharp and very high trap.  

Again finite size effects close to perfect nesting in the infinite system, or half filling, are more or less negligible with only slightly varying critical scales, see the right plot in Fig.~\ref{fig:squarefs}. There, no ambiguity in the resulting phase occurs, an antiferromagnetic divergence is found at all sizes. Even the ordering pattern is not influenced much by varying the system size. This is in agreement with the observations by Mazurenko et.~al.~\cite{mazurenko_cold-atom_2017}.
Upon increasing $t'$ the critical scales are decreasing. The finite size effects become more pronounced with increasing $t'$. At $t'\approx -0.26$ we find a transition to a pairing divergence for the two largest systems. For these two, the critical scales are comparable to the ones in the thermodynamic limit. We only encounter superconductivity for a radius larger than $r=20$, additionally the phase transition between AFM and SC is shifted towards higher values of $-\frac{t'}{t}$. This already hints at possible complications for the experimental setups. In the round case we are not able to reach the thermodynamic limit, even for $r=30$, as the phase boundaries do not match the ones from the thermodynamic limit even at the largest radius. As for the square lattice case, no pure bulk $d_{x^2-y^2}$ superconductor is recovered, as again in the complete bulk a mixing with the $s$-wave component is present.

\begin{figure}[!htpb]
    \includegraphics[width = 0.49\columnwidth]{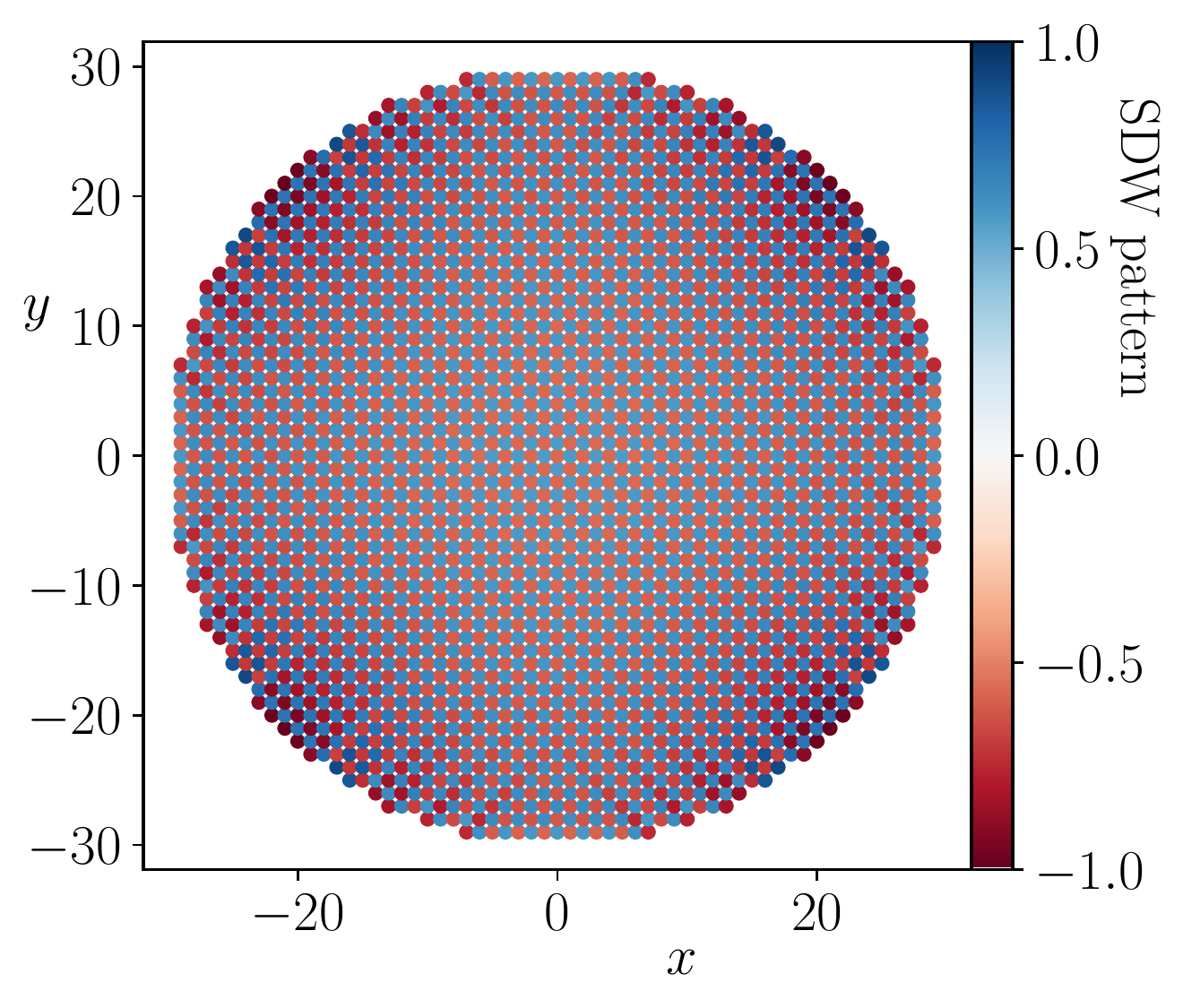}
   \includegraphics[width = 0.49\columnwidth]{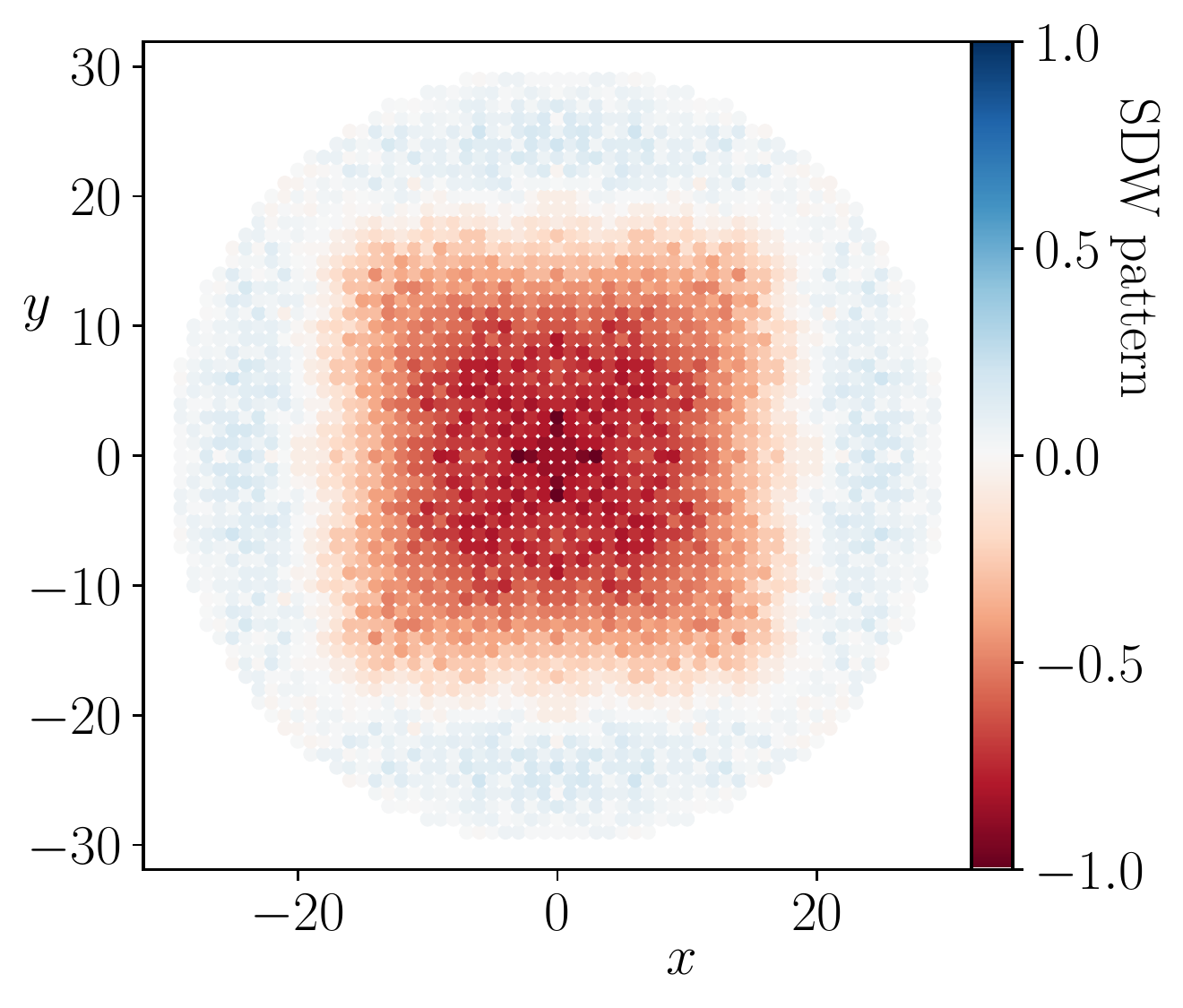}
   \vfill
   \includegraphics[width = 0.49\linewidth]{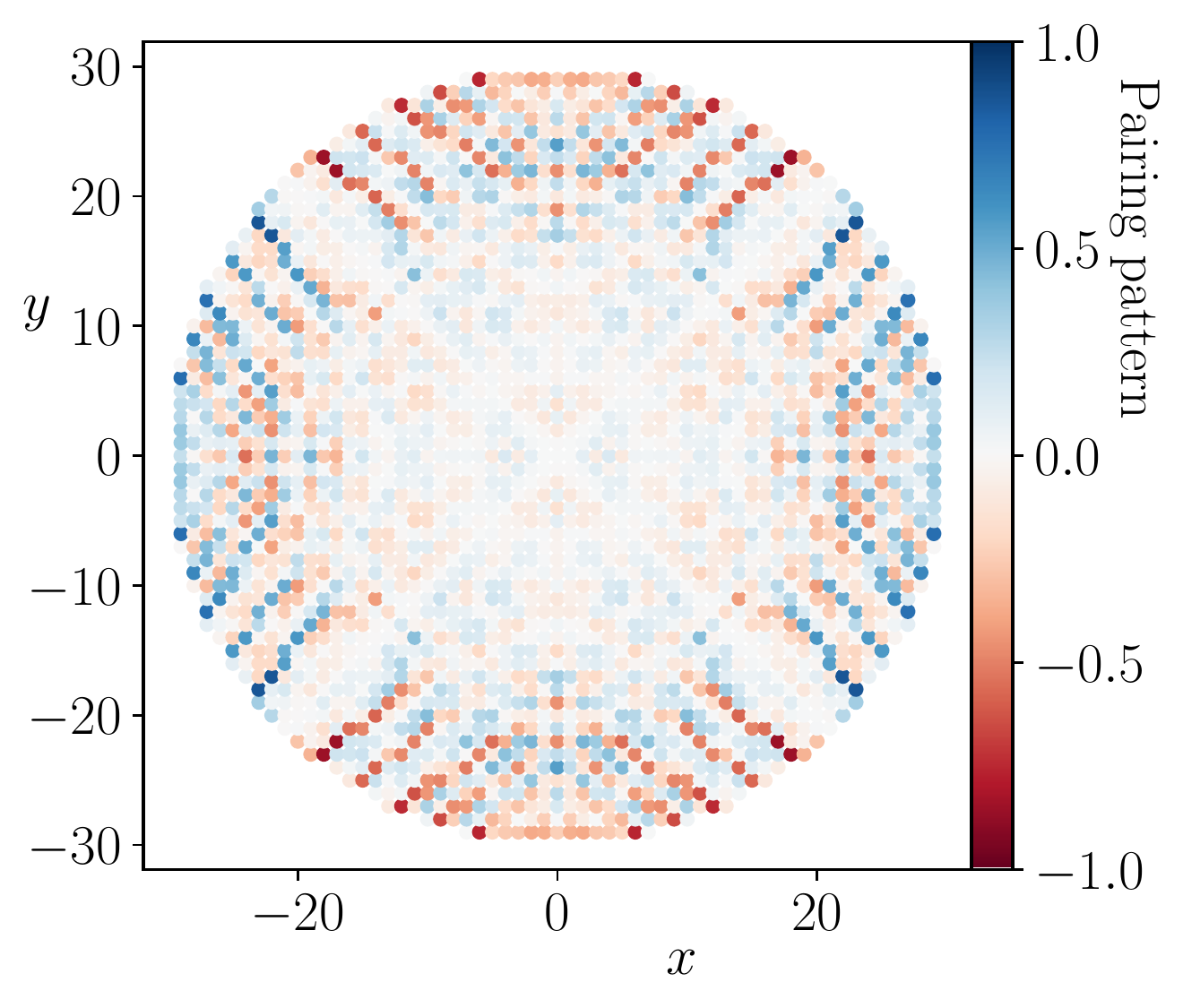}
    \includegraphics[width = 0.49\linewidth]{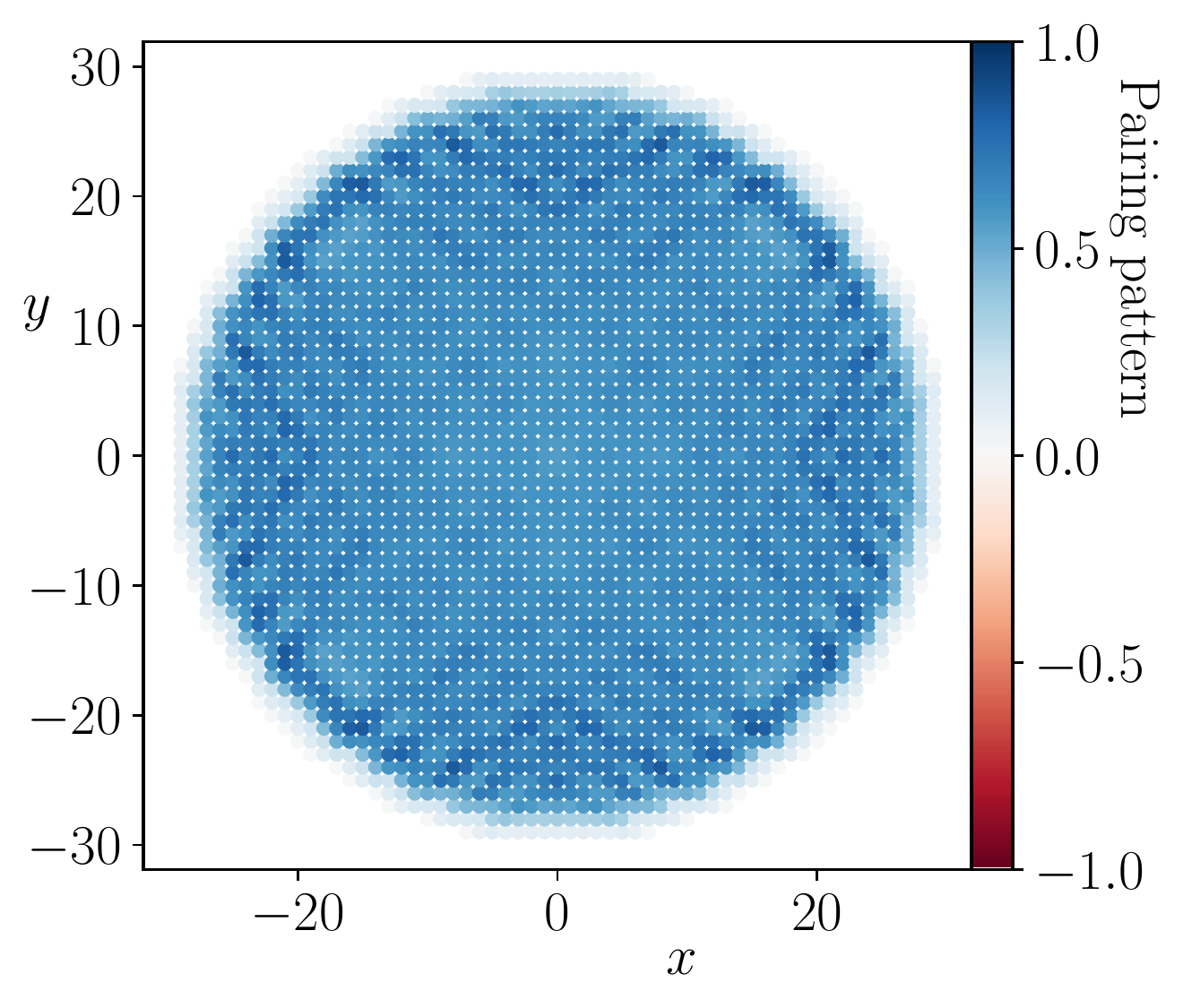}
\caption{Examples for the leading eigenvectors for each phase in the square lattice Hubbard model with circular open boundary conditions and without any confining potential at $r = 30$. Calculations are performed using $U=3$, $T=10^{-3}$ and including nearest-neighbor correlations. In the upper row the leading eigenvectors of the $C$-channel are shown at $t'=0$ in the left, which is an antiferromagnet, and $t'=-0.4$ in the right image, forming a bulk square shaped ferromagnet. The lower row shows the $s$-wave (left) and symmetrized $d_{x^2-y^2}$ (right) contribution of the leading eigenvector of the $P$-channel at $t'=-0.26$. The $P$-channel eigenvector indicates $d+s$ ordering tendencies.}
\label{fig::roundordering}
\end{figure}

We observe a strong size dependence for high $-\frac{t'}{t}$ values where, in the infinite model, the ferromagnetic phase occurs. This can be explained by considering the orderings, see Fig. \ref{fig::roundordering}; Whereas the antiferromagnetic and superconducting phases more or less comply with the circular shape of the lattice (up to boundary effects and the already mentioned mixing), the ferromagnetic phase does not. In fact, the resulting magnetization pattern has a square like shape for the largest size. This of course is not matching to the lattice constraints and therefore ferromagnetic order is suppressed by boundary effects. Before the formation of the square like pattern, it consists of five localized peaks, one in the center and the others at the edges of the square, at increasing sizes they become more and more connected and fuse ultimately to the square-like shape shown. { Changing  from square to circular open boundary conditions significantly deteriorates the convergence of the phase diagram with respect to system size, but the way the orderings change does not differ significantly.}

\subsection{Effects of the trap potential}

\begin{figure}[!htpb]
    \centering
    \includegraphics[width = 0.49\columnwidth]{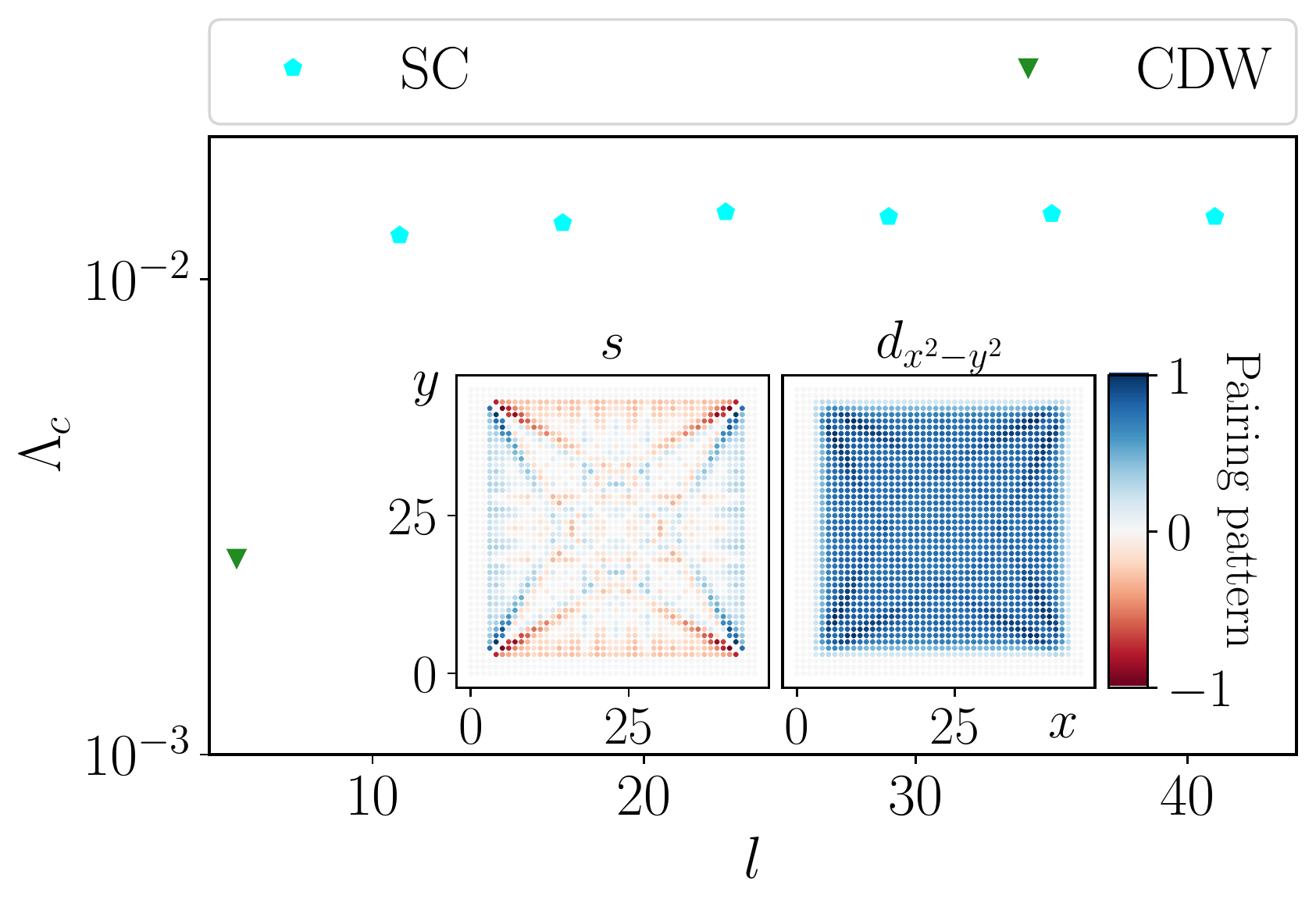}
    \includegraphics[width = 0.49\columnwidth]{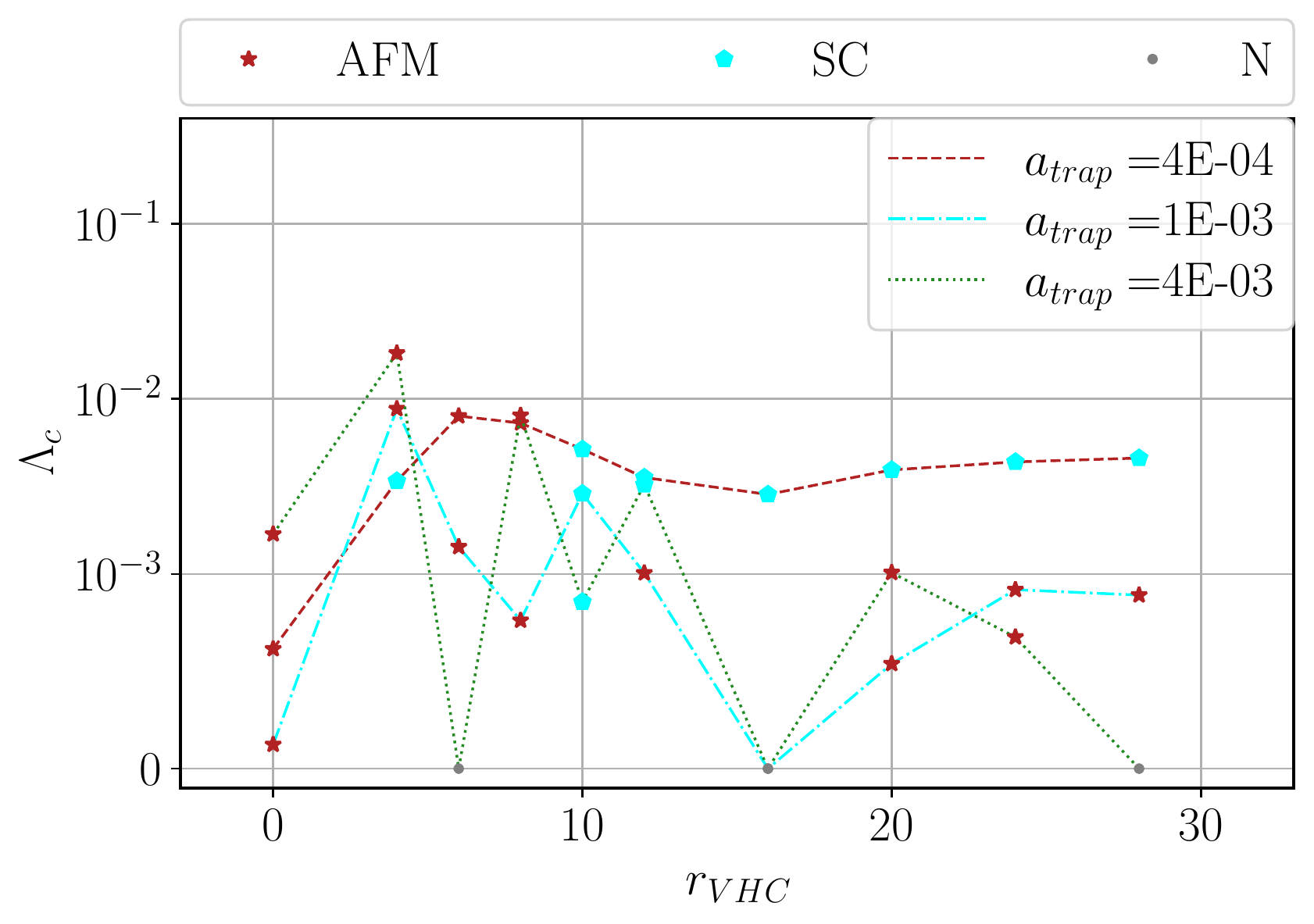}
    \caption{We show the phases of a square lattice Hubbard model with open boundary conditions in which a second box potential with finite height is embedded on the left. Here, the side length $l$ of the inner boxes are given on the $x$-axis. On the right we show the phases of a square lattice Hubbard model with circular open boundary conditions, in which we insert a quadratic trap and vary $r_{VHC}$ at different values of $a_{trap}$. The $y$-axis gives the critical scales, and the shape of the data points encode the resulting phase. In the insets in the left plot, the $s$ and $d_{x^2-y^2}$ component of the leading pairing eigenvector at the largest system size are shown. Calculations are performed using $U=3$, $T=0$ and including nearest-neighbor correlations at $t'=-0.28$. We encounter antiferromagnetism (AFM), charge density waves (CDW), superconductivity (SC) and no divergence (N) at low scales.}
    \label{fig:trap_sqare}
\end{figure}

We now turn to the third setup, a square lattice Hubbard model with open boundary conditions in which a second box potential with finite height is embedded. This is realized by an on-site potential at the two sites closest to the boundary at the upper and the right edge and at the three sites closest to the boundary at the lower and left edge. By this we can verify that the boundaries existing beyond the box potential do not influence our finding. The on-site potential is set higher than the bandwidth, i.~e.~$V^{trap} = 5.0$, suppressing particles in this region. The difference to straight forward open boundaries is that due to the finite height, the electronic wavefunctions are allowed to be non-zero at the boundary. Inside the potential well, the wavefunction decays exponentially. In this case, we are mainly interested in the superconducting phase, thus we choose $-t' = 0.28$. For simplicity we set $T=0$ and apply a sharp frequency cutoff. The results are summarized in the left plot in Fig.~\ref{fig:trap_sqare}.

At the smallest system size, we find a charge divergence. Starting from an effective system length of $l=12$ we observe a $P$-channel divergence. Again the $s$-wave component is dominant at small system sizes and the divergences still show similar patterns as for the open boundary case. But compared to the case without the embedded potential, the $s$-components are less pronounced in the bulk; thus, the confining potential reduced the finite size effects. Thereby, such a potential could reduce the system size necessary to resolve bulk $d_{x^2-y^2}$-superconductivity.  
Such a square box trap potential is of course hard to realize experimentally, therefore we next proceed with a potential which is closer to experiments. 

In experiments, the lattice is generated by a modulated square potential~\cite{tarruell_quantum_2018}, therefore the fourth case investigated is a square lattice Hubbard model with circular open boundary conditions, in which we insert a quadratic trap as on-site potential:
\begin{equation}
    H = H_0 + \sum_{i, \sigma} a_{trap} |\vec{r}_i|^2c^{\dagger}_{i,\sigma} c_{i,\sigma}.
\end{equation}
The coordinate system is arranged such that its origin coincides with the center of the lattice. The trapping is embedded in a round lattice with $r = 30$ and the temperature is fixed to $T=10^{-3}$.
The potential is at first fixed at the center, meaning that the Van-Hove condition is only fulfilled there. We now vary the trap curvature $a_{trap}$ to examine whether superconductivity can still be observed.
\begin{figure}[!htpb]
    \centering
    \includegraphics[width = 0.49\columnwidth]{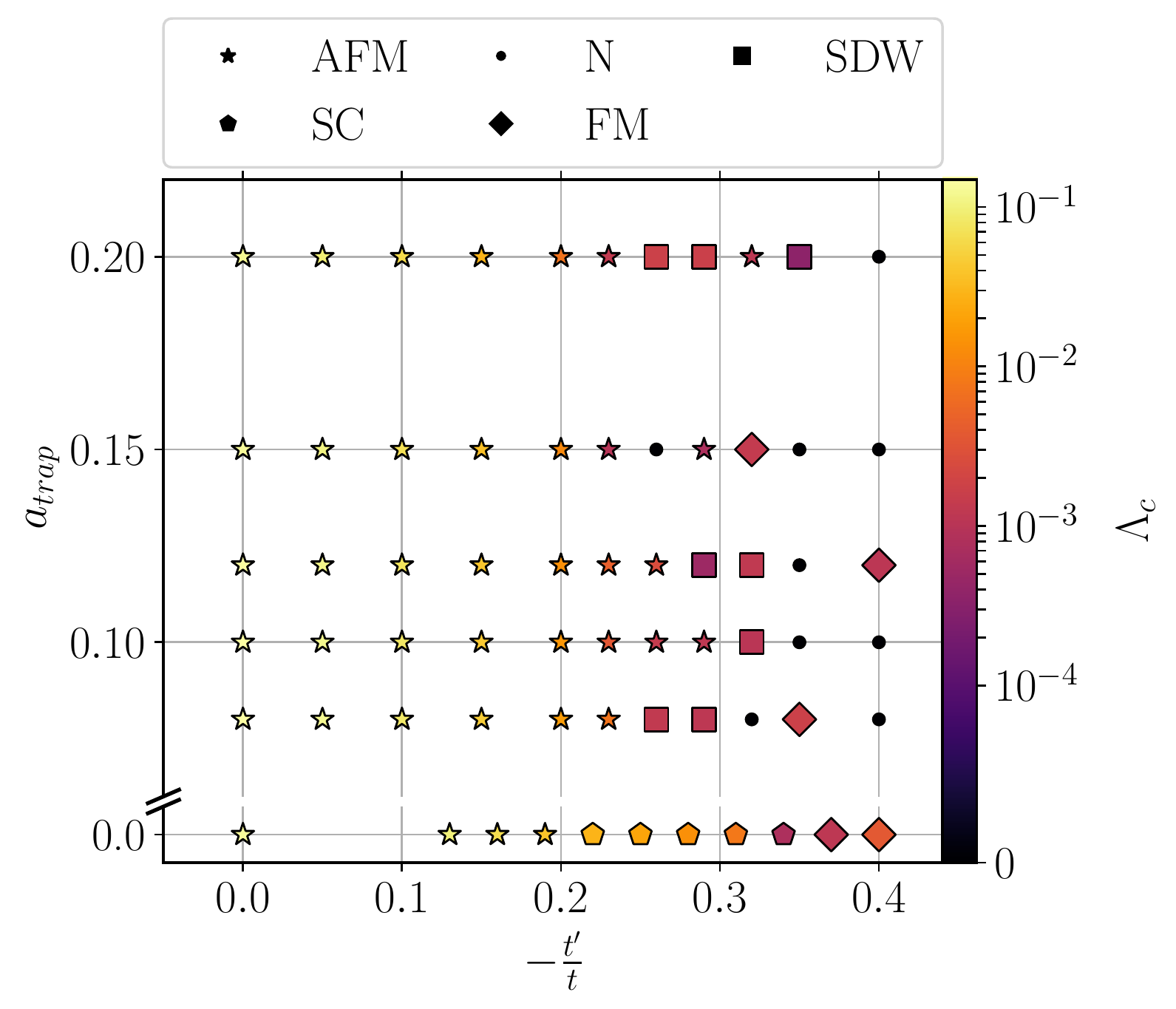}
    \includegraphics[width = 0.49\columnwidth]{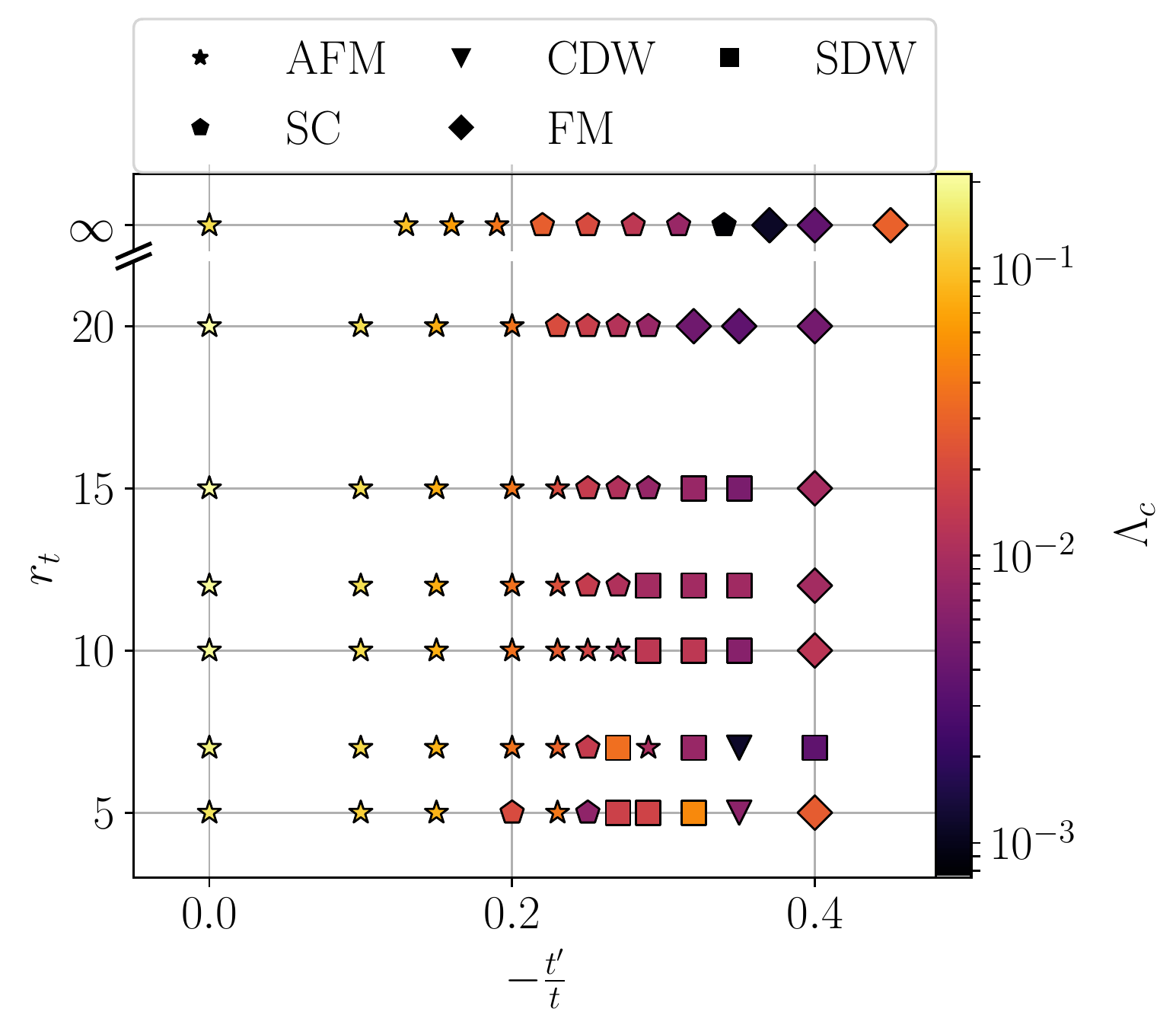}
    \caption{The phase diagram of a square lattice Hubbard model with circular open boundary conditions with an embedded quadratic trap is shown on the left and the phase diagram of a square lattice Hubbard model with circular open boundary conditions with embedded experimental trap is shown on the right, both at varying nearest-neighbor hoppings $\frac{t'}{t}$. Calculations are performed using $U=3$ and including nearest-neighbor correlations at $T=10^{-3}$ and $r=30$ for the left plot and $T=0$ and $r=35$ for the right plot. The y-axis gives the curvature of the trapping potential $a_{trap}$ in the left and the radius of the constant particle number disc in the right plot. The critical scales $\Lambda_c$ are encoded in the color scheme, whereas the type of the diverging phase is given by the shape of the data point.  The thermodynamic limit results are given as a reference in the lower/upper parts of the plots. We encounter antiferromagnetism (AFM), general spin density waves (SDW), ferromagnetism (FM), charge density waves (CDW), superconductivity (SC) and no divergence (N) at low scales. }
    \label{fig:trap_round_nofix}
\end{figure}

At low values of $-\frac{t'}{t}$ the critical scales do not differ much, see the left plot in Fig.~\ref{fig:trap_round_nofix}. Additionally, the emergent phase is matching the expectation. With lowering $a_{trap}$ the critical scale converges to the same values observed in Fig.~\ref{fig:trap_sqare} for the same system size. The phase diagram is cut-off by the temperature for high $-\frac{t'}{t}$ values. We observe that the parameter regime in which we expect superconductivity is dominated by spin-wave ordering, {which is in line with earlier  investigations~\cite{cone_optimized_2012}}. The superconducting phase vanishes due to the confinement of the effective system size in which the Van-Hove condition (VHC) is fulfilled. The emergent spin divergences are pinned to only a very small region close to the center of the lattice, the VHC fulfilling region for which $\mu_i+V_i \approx 4t'$. As we cannot follow the RG flow into the symmetry broken region, we cannot resolve possibly coexisting states at lower energy scales outside of the VHC fulfilling region. All in all, as in experiments $a_{trap}$ cannot be reduced arbitrarily, it is unlikely to observe bulk $d$-wave superconductivity in this simple setup. More sophisticated methods of trapping the ultracold atomic gas must be applied. We did not recover superconductivity for any reasonable trapping curvature. In contrast, enlarging the VHC fulfilling region by shifting the minima to a ring, leads to a recovery of the superconducting phase which is then bound to this region. Therefore, this offers a possible way to manipulate the spatial dependence of the order parameter, e.g.~restricting superconducting order to a ring.
{ Here we observed first, that a simple trap smoothens the open boundary conditions, reducing the finite size effects and second that with vanishing size of the VHC-fulfilling region the superconducting phase vanishes too.}

\subsection{Trap-shaped superconductivity}

To examine the possibility of shaping the ordering by changing the trapping potential, we change the radius of the region fulfilling the VHC by changing the potential to $a_{trap}(r^2-r_{VHC}^2)$, thereby the VHC fulfilling region is now a ring of radius $r=r_{VHC}$ and not a point at $r = 0$ anymore. The slope at the zero crossing is increasing with increasing $r_{VHC}$, such that the width of the ring in which the VHC approximately holds shrinks to zero for large radii. The two effects counteract each other such that an optimal value for $r_{VHC}$ to promote superconductivity exists. { The emergence of ring like pairing correlations were observed earlier~\cite{chiesa_magnetism_2011} in a DQMC study, we now want to study its radius dependence and stability w.r.t.~variations of the trapping curvature $a_{trap}$.}
We choose $T=0$ combined with a sharp cutoff in order to enlarge the system size to $r =35$. We apply three different trapping curvatures, $a_{trap} = 4\cdot 10^{-4}, 1\cdot 10^{-3} \text{ and } 4\cdot 10^{-3}$ (note that the smallest value is still one order of magnitude too large to recover the $d$-wave superconducting phase in the simple setup above at a next-nearest-neighbor hopping $t' = -0.28$, where we expect superconductivity). We vary the VHC fulfilling radius $r_{VHC}$ and track the leading divergences of the channels and their changes. 
\begin{figure}[!htpb]
    \includegraphics[width = 0.49\columnwidth]{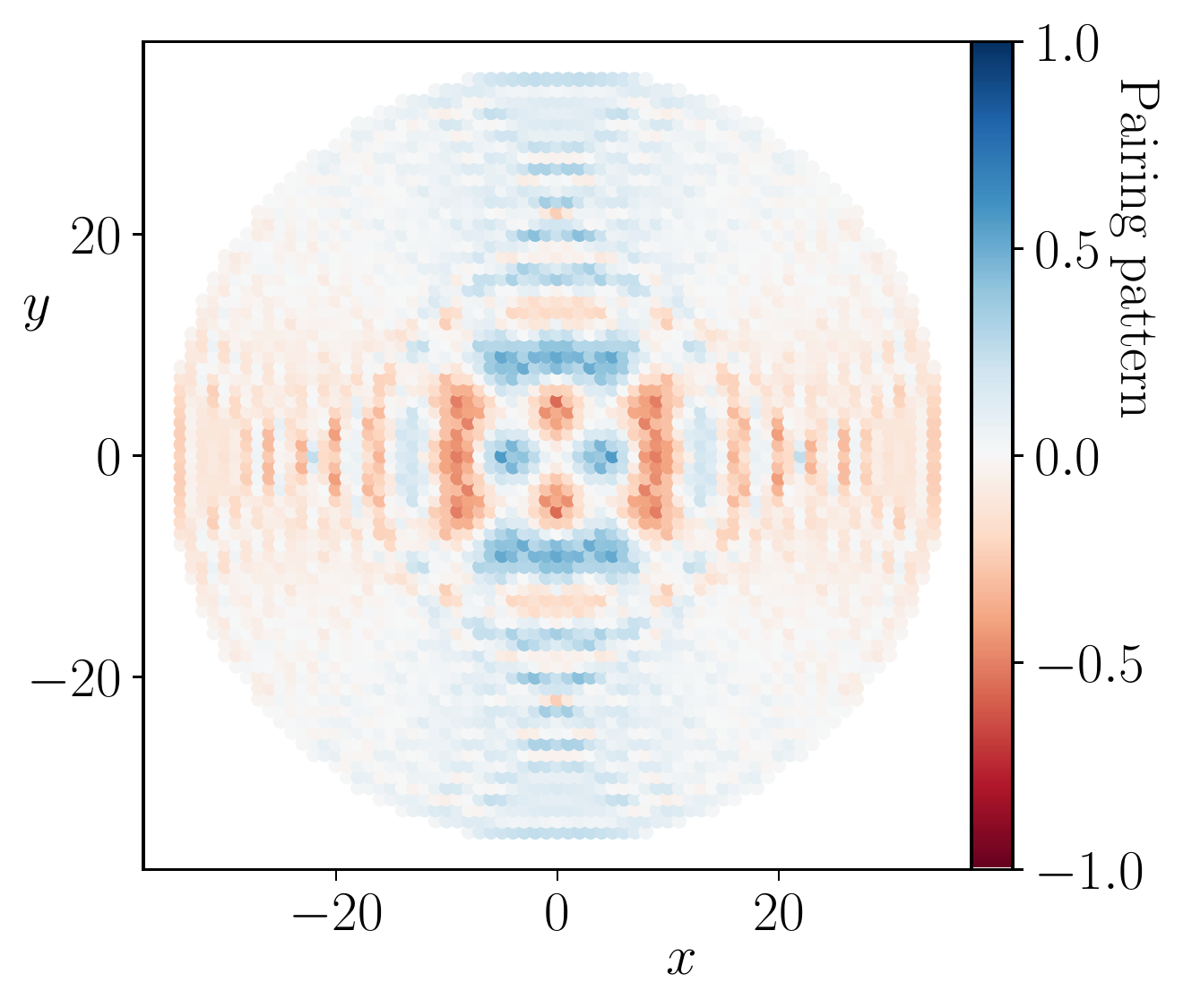}
   \includegraphics[width = 0.49\columnwidth]{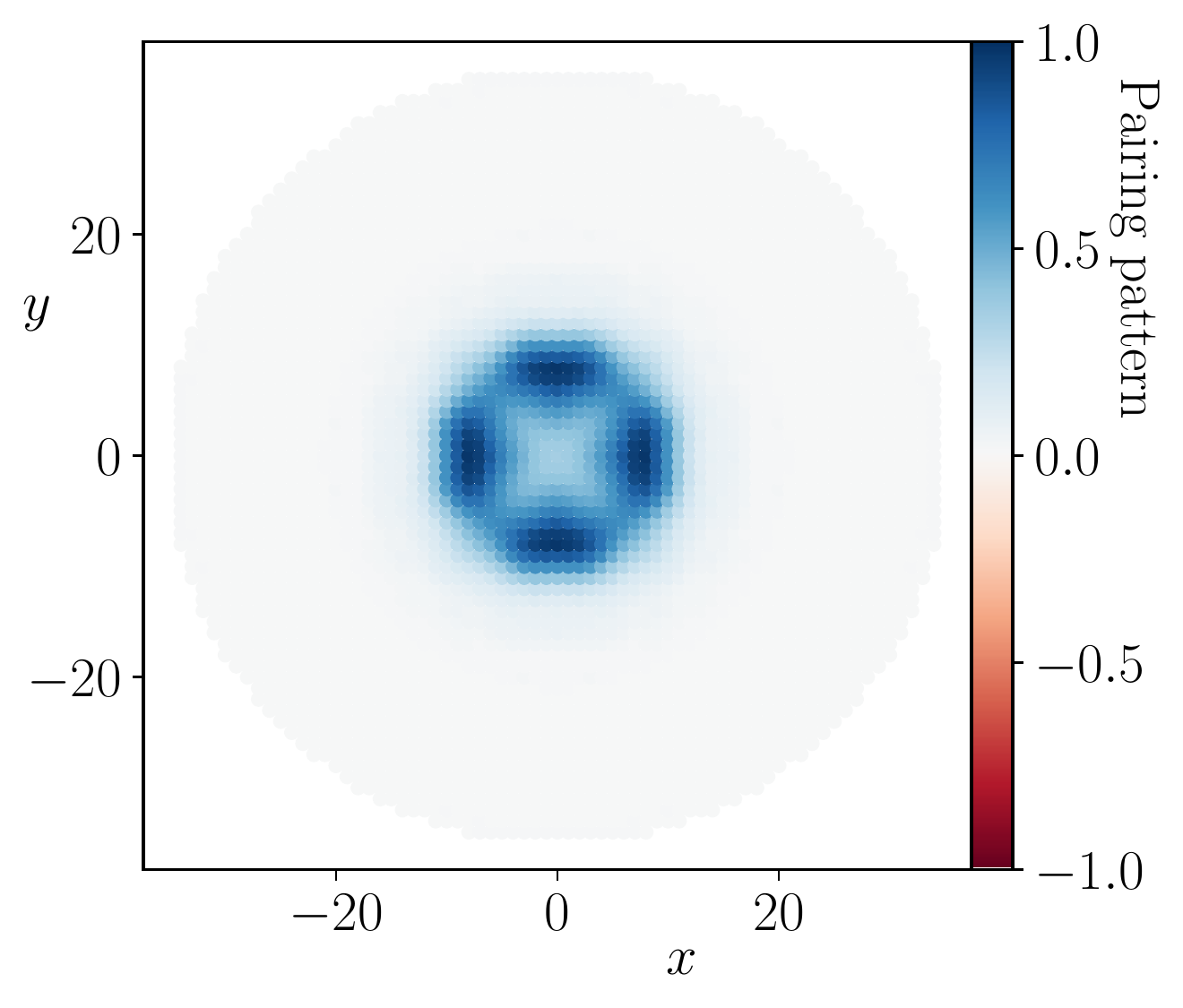}
   \vfill
   \includegraphics[width = 0.49\linewidth]{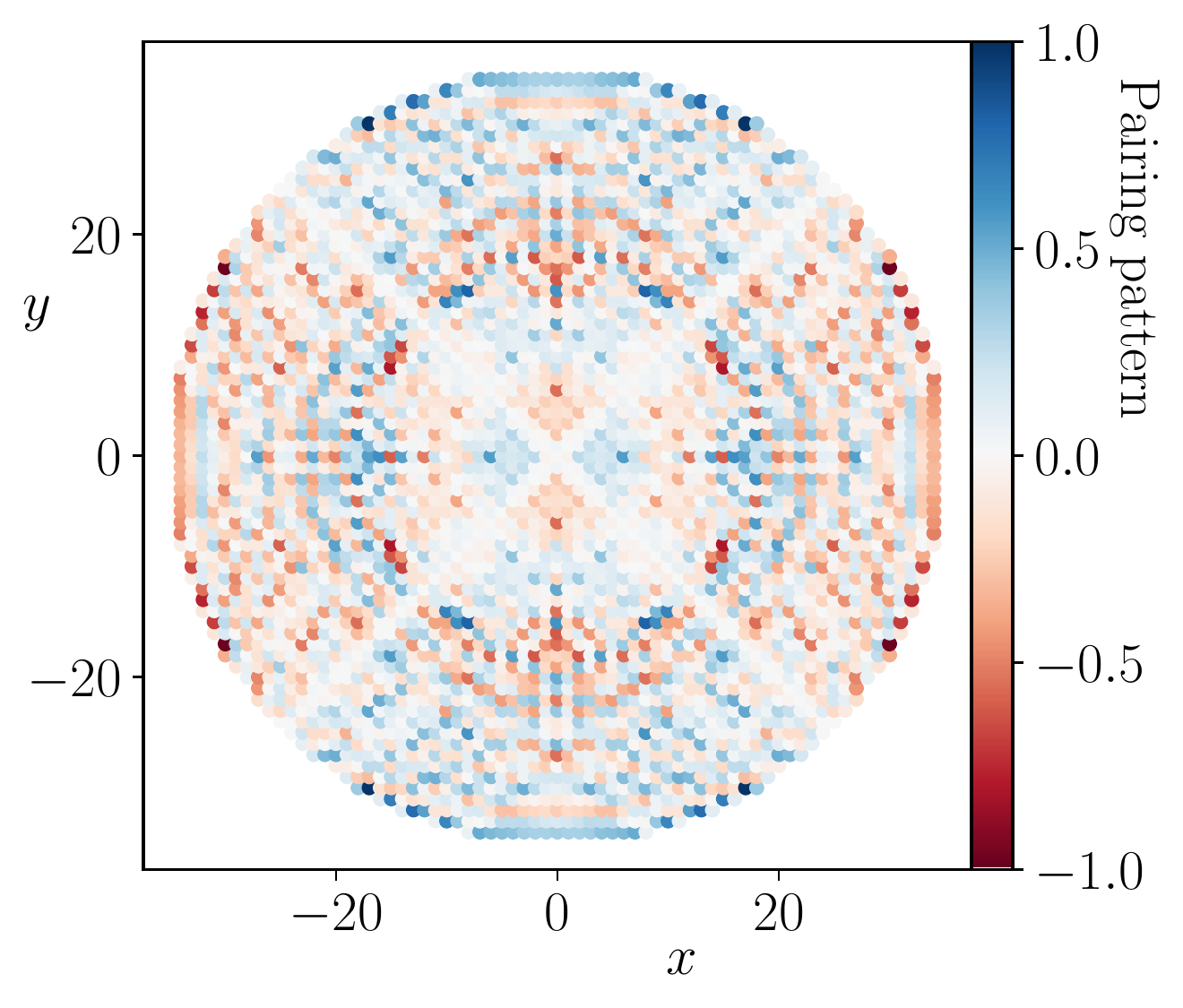}
    \includegraphics[width = 0.49\linewidth]{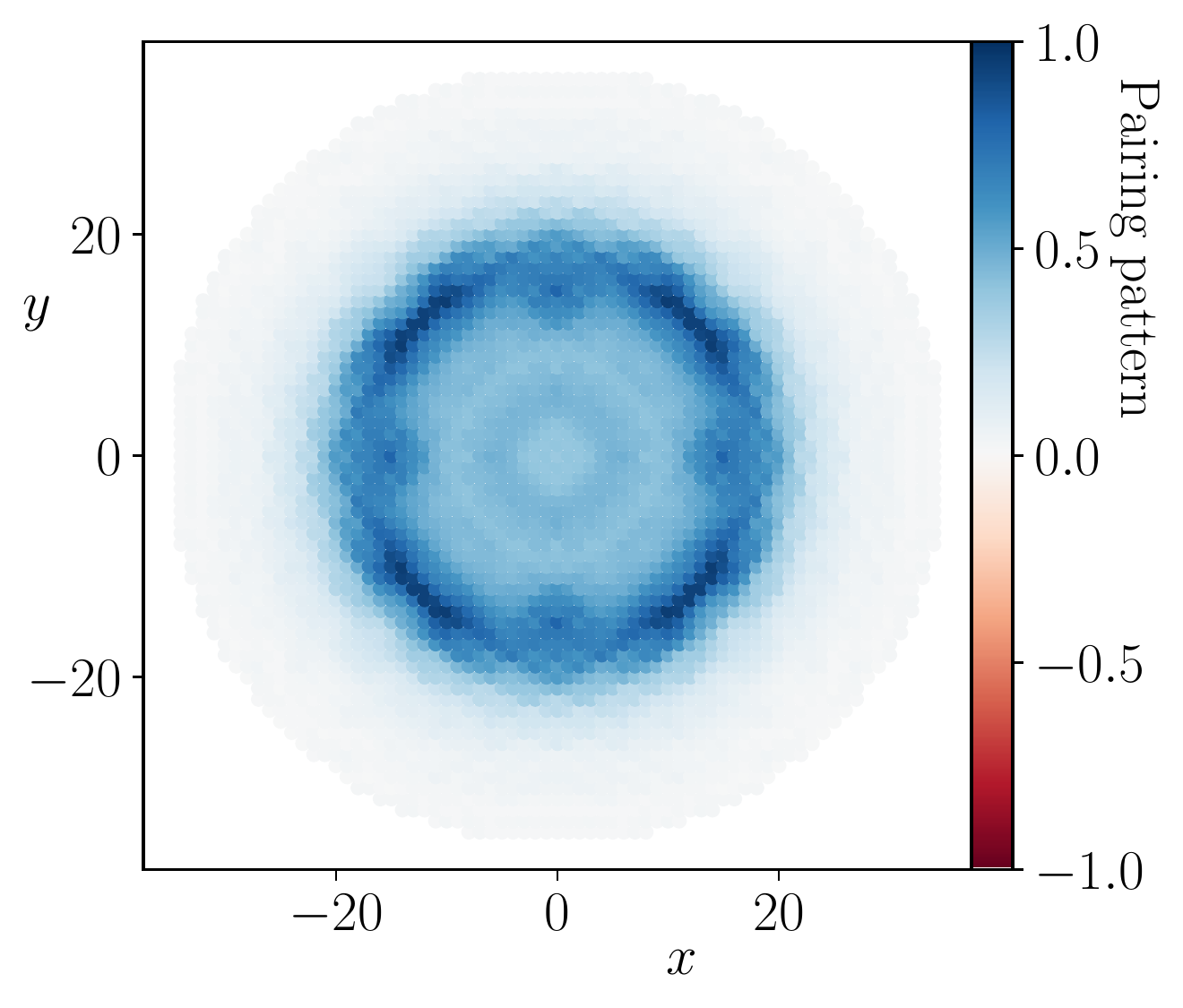}
\caption{Superconducting ordering parameters for two different radii of the $VHC$-fulfilling region at $U=3$, $T=0$ and $t'=-0.28$. The upper row shows $r_{VHC} = 10$ at $a_{trap} = 10^{-3}$ and the lower row $r_{VHC} = 20$ at $a_{trap} = 4\cdot10^{-4}$, in the left column the $s$-wave contribution of the largest eigenvector is shown and in the right column $d_{x^2-y^2}$ contribution is shown.}
\label{Fig::SC_shaped}
\end{figure}

For the smallest curvature, we find three phase transitions upon increasing the VHC radius $r_{VHC}$, see right plot in Fig.~\ref{fig:trap_sqare}. The first occurs between $r_{VHC}=0$ and $r_{VHC}=4$. At $r_{VHC}=0$ we obtain a localized spin-divergence of the vertex, which hints at a localized state emerging due to the potential, similar to what has been observed in~\cite{chanda_coexistence_2020}. At $r_{VHC}=4$ we find a $d_{x^2-y^2}$-divergence which vanishes again upon increasing $r_{VHC}$ further. Between $r_{VHC}=4$ and $r_{VHC}=10$ we find an antiferromagnet. If we further increase the radius we obtain a superconductor, which at first has mainly a $d_{x^2-y^2}$ component, and is circular shaped with no clear hole in the center, similar to the upper plots in Fig.~\ref{Fig::SC_shaped}. At larger radii, the pairing amplitude is reduced in the center and has main weight on a thin ring at the VHC-radius, this leads to a slight reduction of critical scales. Additionally, the s-wave component is increasing due to the weaker screening of the boundaries, see the lower plots in Fig.~\ref{Fig::SC_shaped}. At the largest radii, the s-wave component dominates due to the proximity to the boundary, which leads to a slight increase of the critical scale. All these findings underline the importance of the radius and the shape of the VHC fulfilling region. We did not observe a transition to a disconnected ring, which would require a hole in the gap amplitude at the center. Such a pure ring has a different topology and might host exotic topological quantum states. Here, the dependence of the critical scale on $r_{VHC}$ is rather smooth, in contrast to the two larger curvatures. In those systems we mostly obtain local divergences bound to the VHC fulfilling region. At $r_{VHC} = 10$ we find a superconducting divergence in all three setups, where the weight of the $s$-wave component compared to the $d_{x^2-y^2}$-component is increasing with the trapping curvature.

Remarkably the smooth boundary conditions seem to suppress effects arising due to open boundaries, see for example the small $s$-wave component for $r_{VHC}=10$ in Fig.~\ref{Fig::SC_shaped}. Thus it could become easier to observe a bulk $d_{x^2-y^2}$-superconductors with a smooth trap than with a sharp trap. Additionally, we were able to show that we can shape the superconducting phase to a circle. Inspired by this, an experiment could try to shape the superconducting phase by applications of specifically designed traps. This could result in new types of topological superconductivity or designed superconducting currents. For this it would be helpful to create a particle depletion inside the ring, which is achievable by the application of a Mexican-Hat potential $ar^4-br^2 $ (with $a,b>0$) for which the minima is set to fulfill the VHC. To this end, further investigation is needed either numerically or experimentally. { In general, it will be very interesting to investigate the possible uses of trapping potentials for experimental realisations of exotic phases of matter.}

\subsection{Experimental Trap}
The last setup we investigate is a square lattice Hubbard model with circular open boundary conditions, in which a trap reconstructed from experiments is embedded.
The information given by Mazurenko et.~al.~\cite{mazurenko_cold-atom_2017} allows for an approximate modelling of their trapping potential. It consists of the square lattice part, which we discussed in detail above, and an additional digital micromirror device (DMD) potential to create a disk of constant number density. We use a piece-wise defined function consisting of a constant disk, a Gaussian which emulates the DMD and a square potential to capture the lattice potential, see Eq.~(\ref{Eq:picewise_trap}). The parameters are chosen such that the particle density at half filling as a function of the distance to the center is roughly matching the one given in  \cite{mazurenko_cold-atom_2017}. 

\begin{equation}
a(x) = \left\{
\begin{array}{ll}
0  &\text{if $r<r_t$} \\
\gamma  \left(e^{\frac{(r-r_t-\Delta)^2}{2\sigma_t^2}} - e^{\frac{\Delta^2}{2\sigma_t^2}}\right)  &\text{if $r_t<r<r_t+\Delta$} \\
\gamma \left(1-e^{\frac{\Delta^2}{2\sigma_t^2}}\right) - b\cdot(r-r_t-\Delta)^2  &\text{if  $r>r_t+\Delta$}
\end{array}
\right. 
\label{Eq:picewise_trap}
\end{equation}

In total we have five free parameters, the gaussian prefactor $\gamma$, the radius of the disc $r_t$, the gaussian regime parameter $\Delta$, the width of the gaussian $\sigma_t$, and the quadratic prefactor $b$. The trap is designed continuous but has kinks. In the following we vary the trap radius and choose
\begin{equation*}
\gamma = 85, \quad \sigma_t = 20, \quad \Delta = 5, \quad b = 0.01666.
\end{equation*}
The VHC is chosen to be fulfilled at all sites within the trap radius.
The difference to the radius variation in the open boundary system lies in the smooth boundaries we obtain due to the Gaussian. Thus, we expect different finite size effects. For example, we expect a less pronounced s-wave component in the superconducting phase due to what was observed in the setups investigated above.

\begin{figure}[!htpb]
\centering
\includegraphics[width = 0.49\linewidth]{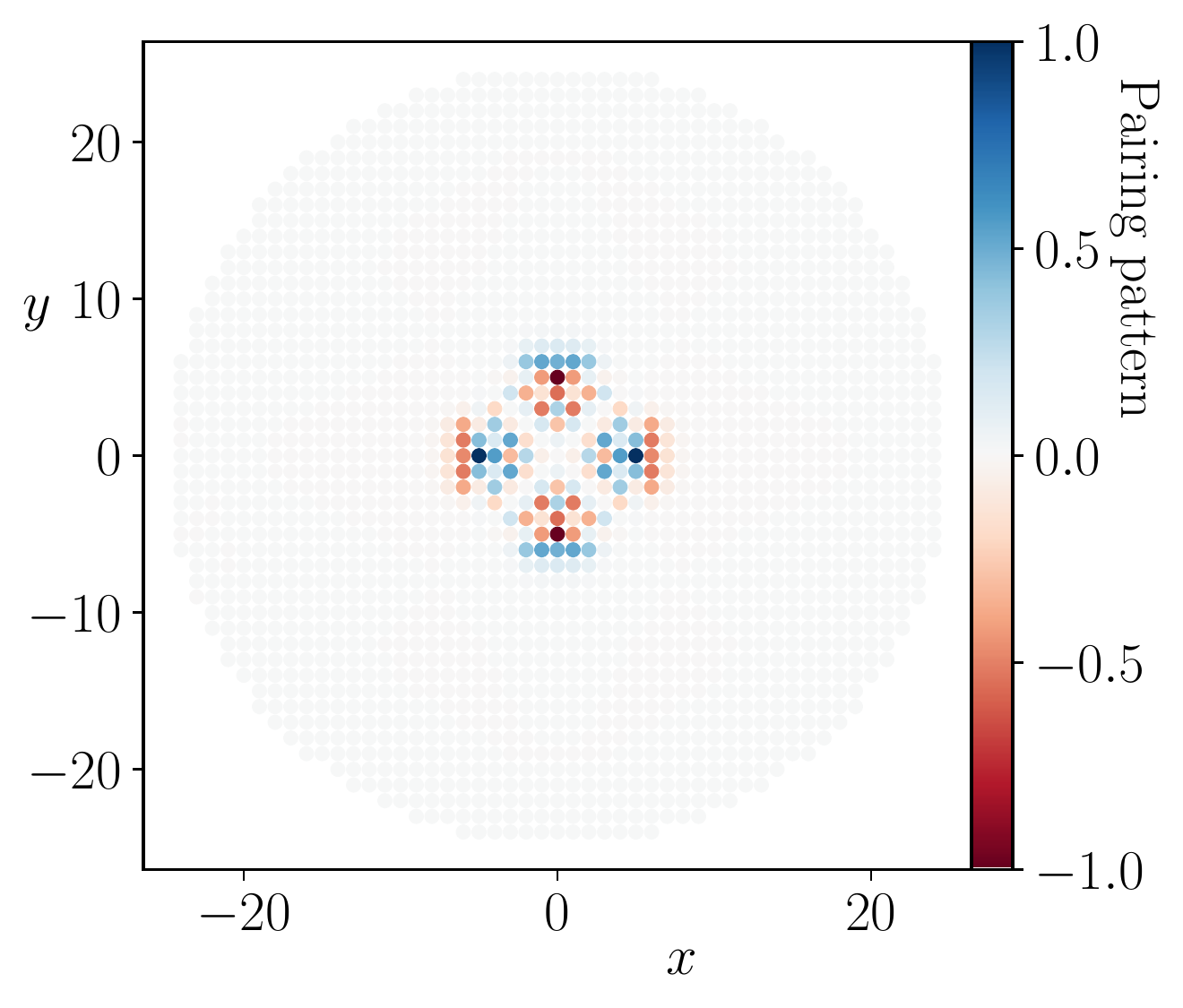}
\hfill
\includegraphics[width = 0.49\linewidth]{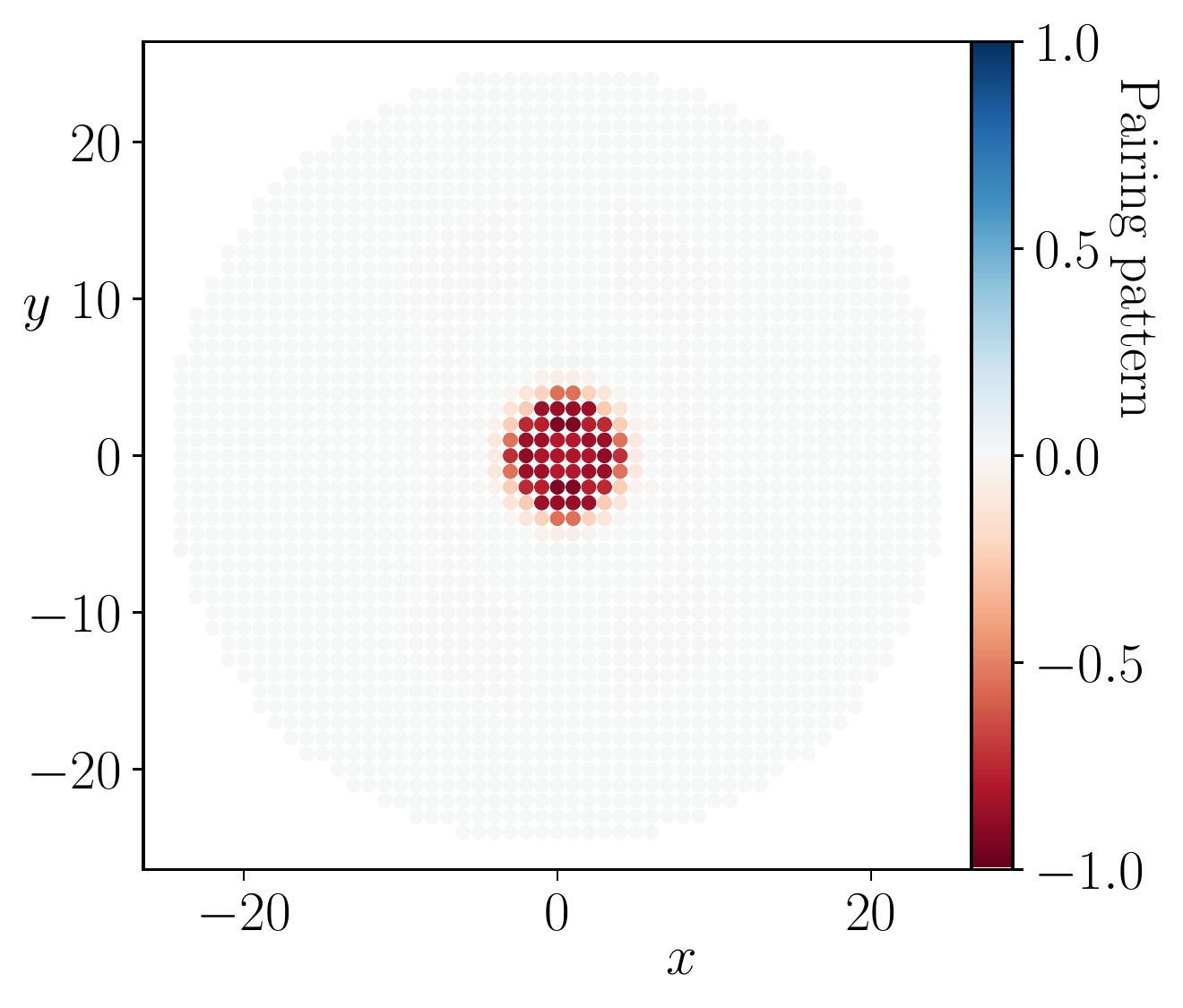}
\vfill
\includegraphics[width = 0.49\linewidth]{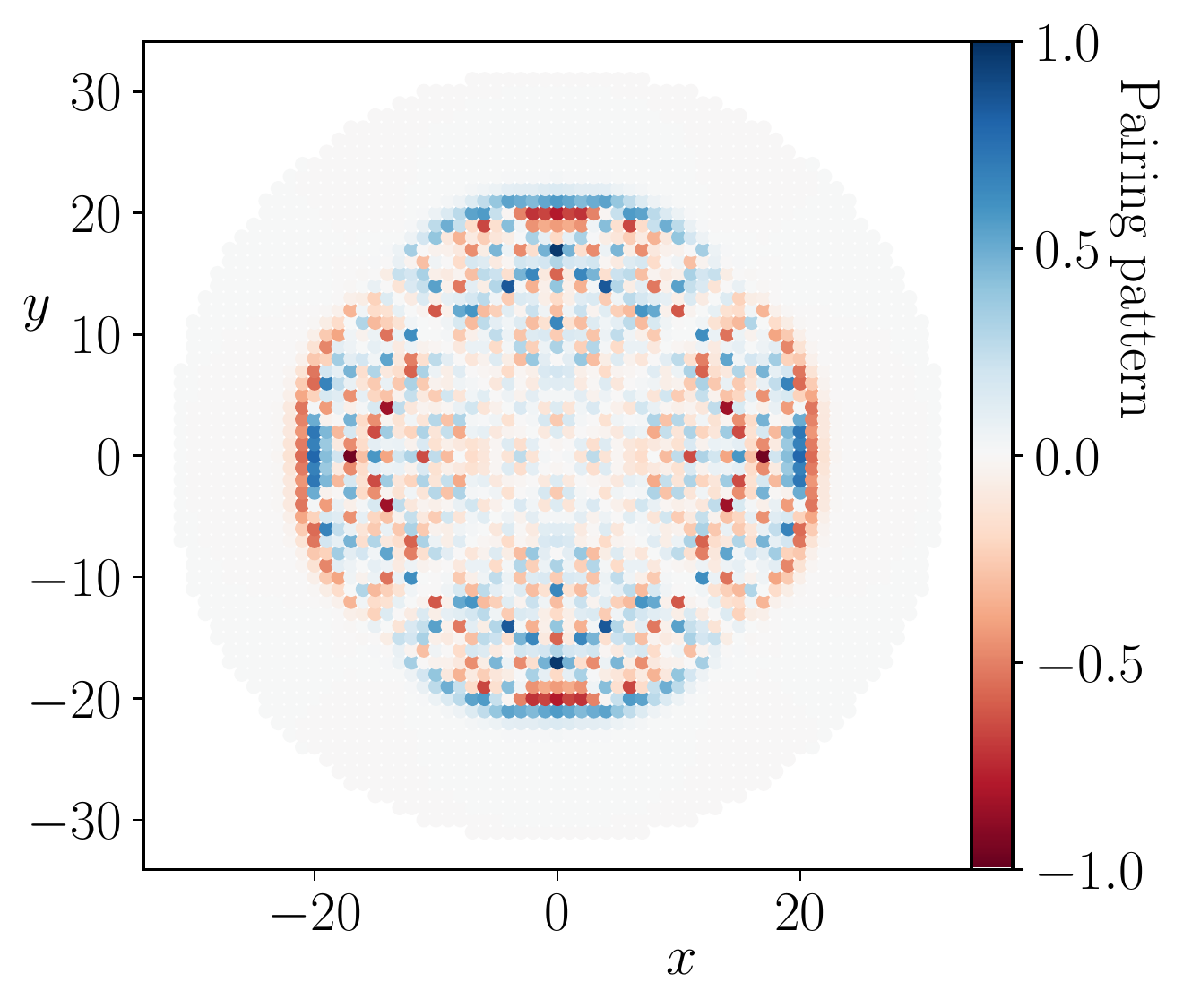}
\hfill
\includegraphics[width = 0.49\linewidth]{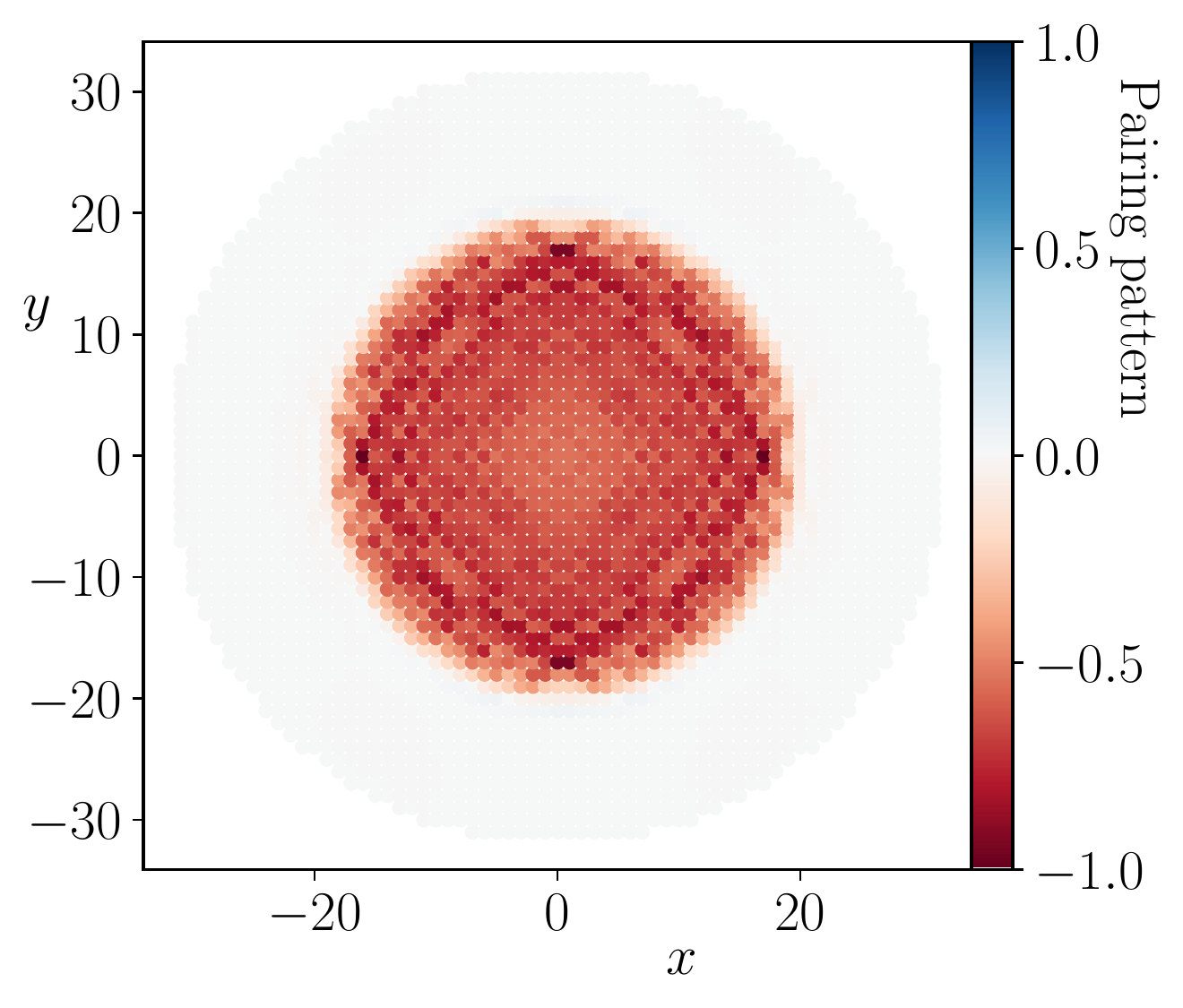}
\caption{Leading $P$-channel eigenvector at $t' = -0.25$ for $r_t = 5$ in the upper, and at $t' = -0.27$ for $r_t = 20$ in the lower row. The left pictures show the on-site component of the eigenvectors whereas the right pictures visualize the symmetrized $d_{x^2-y^2}$ component of it. Calculations are performed using $U=3$, $r=35$, $T=0$ applying a sharp-cutoff and including nearest-neighbor correlations.}
\label{fig::sup_large}
\end{figure}

The finite size effects are again small for weak to intermediate values of $t'$, which is expected in analogy to the results we obtained for the other models {as well as earlier studies employing similar trappings~\cite{cone_optimized_2012}}. The critical scale, and thus the critical temperature, is increasing slightly upon increasing the lattice size. At $r=20$ and $t'=-0.1$ we observe that the critical scale is approaching the known FRG critical scale of $\Lambda_c = 0.11$ at $U=3$~\cite{lichtenstein_functional_2018}.  For increasing values of $t'$ the phases are less stable. We encounter a superconducting divergence at the two smallest system sizes, but they have support on only a few sites, see Fig.~\ref{fig::sup_large}. The occurrence of these $P$-channel divergences seems to be a fine-tuned problem as increasing $t'$ or the radius can make them vanish and reappear. In the $r_t = 10$ model, there is no superconducting phase transition observed.

Still, the superconducting phases at the lowest system sizes indeed show the expected ordering pattern, namely $d+s$. At higher radii, the $s$-component is again more suppressed in the bulk. The ferromagnetic phase is hampered most by the potential and is not fully recovered at the largest system size studied. It does not fill the whole trap region but manifests itself on a square shaped subsection, this behavior is found in all cases where we applied non-periodic boundary conditions. Thus, it should be observable in experiments. We observe that the trapping reduces the effects of the open boundary conditions especially concerning the $s$-wave component of the superconducting phase.  At the largest radius, we obtain a similar distribution of phases in the $t'$ domain as for the thermodynamic limit case~\cite{lichtenstein_high-performance_2017}, with the exception that the transition between superconductivity and the ferromagnetic phase is shifted to lower values of $-\frac{t'}{t}$.

The question arises whether a pure bulk $d_{x^2-y^2}$ superconductivity is observable at the largest system size. In Fig.~\ref{fig::sup_large} the superconducting order parameters are visualized at $t'=-0.23$ and $r_t =20$. We observe that the $s$-wave component of the eigenvector is vanishing in a small region at the center, such that there, we obtain a local bulk $d_{x^2-y^2}$ superconductor. With such a setup we thus might be able to recover the thermodynamic limit in the laboratory at reasonable effort. { Comparing the present setup to the simpler trap shapes studied above, we observe that the two ingredients most important for reaching a fast convergence in system size in the studied Hubbard model are large sizes of the VHC fulfilling regions in combination with smooth trapping potentials. An optimal design of the latter will certainly help to reduce the effective system size required in the future.}

\section{Conclusions}

We studied the finite-size effects in round- and square-shaped fermionic Hubbard models on an underlying square lattice, as well as the influences of three different trapping potentials on the (finite-size) phase diagram of the $2$D Hubbard model. We found that the application of open boundary conditions leads to a suppression of the ferromagnetic region of the phase diagram which occurs for larger values of the next-nearest-neighbor hopping amplitude. The finite-size corrections for the weak next-nearest-neighbor hopping are much smaller, which is in line with experiment. The superconducting phase is suppressed at small system sizes and has in general an $d+s$ form induced by the open boundary conditions, with the somewhat unexpected admixture of onsite $s$-wave paring to dominant $d$-wave nearest-neighbor pairing. For the square lattices, the superconducting phase is less suppressed. We observed that designing a specific shape of the trap can tune the divergence encountered, i.~e., we were able to create disk like superconductors with main weight at the outer circle by choosing the Van-Hove condition fulfilling radius appropriately. 
Applying the trapping potential reconstructed from the experiments we observed that the smoothing of the boundaries leads to a smaller suppression of superconductivity. Thus, we observed a stable transition from antiferromagnetism to superconductivity already at $r=15$. The $s$-wave component is observed to be less pronounced compared to standard open boundary conditions, thus it is likely that the equivalence between such an experimental system and periodic boundary conditions model is recovered earlier. 

To obtain a more complete description of the trapped Hubbard models, a next step is to incorporate the disorder introduced by the DMD which was neglected here but could lead to further localization effects.
The next steps from the theoretical point of view are the inclusion of self-energy feedback as well as a single frequency dependence of the vertex~\cite{reckling_approximating_2018,bauer_functional_2014,weidinger_functional_2017}. This would allow for predictions correct up to $U^3$. Especially the effects of the self-energy in open boundary systems could play a crucial role. Predictions for the realistic cuprates interaction strength $\frac{U}{W}\propto1$ are not reliable with the real-space TUFRG due to its perturbative motivation. This issue could be resolved by the combination of dynamic mean-field theory with real-space TUFRG~\cite{wentzell_correlated_2015,vilardi_antiferromagnetic_2019,katanin_extended_2019,taranto_infinite_2014}. 

Our study illustrates that on the experimental side, systems sizes need to be sufficiently large, in addition to reaching a lower temperature. 
However, we illustrate that the trapping should not only be understood as a challenge but can also provide an opportunity to freely shape emergent phases to our needs.

\ack
The authors acknowledge the input on how to write and optimize a high-performance code by E.~di Napoli, D.~Rohe and S.~Achilles. We thank J.\ Ehrlich, L.\ Klebl, N.\ Caci and J.\ Beyer for fruitful discussions.
The Deutsche Forschungsgemeinschaft (DFG, German Research Foundation) is acknowledged for support through RTG 1995 and under Germany's Excellence Strategy - Cluster of Excellence Matter and Light for Quantum Computing (ML4Q) EXC 2004/1 - 390534769. We acknowledge support from the Max Planck-New York City Center for Non-Equilibrium Quantum Phenomena.
Simulations were performed with computing resources granted by RWTH Aachen University under project rwth0514.

\section*{References}
\bibliography{paper_qsim}
\end{document}